\documentclass[prd,superscriptaddress,amsfonts,amssymb,amsmath,showpacs,twocolumn]{revtex4-2}
\usepackage{bm}
\usepackage{amsfonts}
\usepackage{latexsym}
\usepackage{graphicx}
\usepackage{amsmath}
\usepackage{palatino}
\usepackage{mathpazo}
\usepackage{textcomp}
\usepackage{float}
\usepackage{booktabs}
\usepackage{dcolumn}
\usepackage{booktabs}
\usepackage{multirow}
\usepackage{hyperref}
\hypersetup{colorlinks,citecolor=blue}
\usepackage{amsmath}
\usepackage{xcolor}
\usepackage{orcidlink}
\usepackage[caption=false]{subfig}
\usepackage{commath}
\def\jnl@style{\it}
\def\aaref@jnl#1{{\jnl@style#1}}

\def\aaref@jnl#1{{\jnl@style#1}}

\def\aj{\aaref@jnl{AJ}}                   
\def\apj{\aaref@jnl{ApJ}}                 
\def\apjl{\aaref@jnl{ApJ}}                
\def\apjs{\aaref@jnl{ApJS}}               
\def\apss{\aaref@jnl{Ap\&SS}}             
\def\aap{\aaref@jnl{A\&A}}                
\def\aapr{\aaref@jnl{A\&A~Rev.}}          
\def\aaps{\aaref@jnl{A\&AS}}              
\def\mnras{\aaref@jnl{Mon.~Not.~Roy.~Astron.~Soc.}}             
\def\prd{\aaref@jnl{Phys.~Rev.~D}}        
\def\prc{\aaref@jnl{Phys.~Rev.~C}}  
\def\prl{\aaref@jnl{Phys.~Rev.~Lett.}}    
\def\qjras{\aaref@jnl{QJRAS}}             
\def\skytel{\aaref@jnl{S\&T}}             
\def\ssr{\aaref@jnl{Space~Sci.~Rev.}}     
\def\zap{\aaref@jnl{ZAp}}                 
\def\nat{\aaref@jnl{Nature}}              
\def\aplett{\aaref@jnl{Astrophys.~Lett.}} 
\def\apspr{\aaref@jnl{Astrophys.~Space~Phys.~Res.}} 
\def\physrep{\aaref@jnl{Phys.~Rep.}}      
\def\physscr{\aaref@jnl{Phys.~Scr}}       
\def\commat{\aaref@jnl{Comm.~Math.~Phys.}}              
\def\science{\aaref@jnl{Science}}               
\def\cqg{\aaref@jnl{Classical Quant.~Grav.}}            
\def\jpcs{\aaref@jnl{JPCS}}                                     
\def\ijmpd{\aaref@jnl{Int.~J.~Mod.~Phys.~D}}                    
\def\grg{\aaref@jnl{Gen.~Relat.~Gravit.}}               
\def\rpp{\aaref@jnl{Rep.~Prog.~Phys.}}          
\def\npa{\aaref@jnl{Nucl.~Phys.~A}}        
\def\lrr{\aaref@jnl{Living Rev.~Rel.}}                   
\def\jcap{\aaref@jnl{J.~Cosmology Astropart.~Phys.}}    
\def\rmp{\aaref@jnl{Rev.~Mod.~Phys.}}   
\def\epjc{\aaref@jnl{Eur.~Phys.~J.~C}}

\allowdisplaybreaks[1]

\begin{document}

\color{black}       

\title{Constrained $f(Q,T)$ gravity accelerating cosmological model and its dynamical system analysis}

\author{S.A. Narawade\orcidlink{0000-0002-8739-7412}}
\email[Email: ]{shubhamn2616@gmail.com}
\affiliation{Department of Mathematics, Birla Institute of Technology and
Science-Pilani,\\ Hyderabad Campus, Hyderabad-500078, India.}

\author{M. Koussour\orcidlink{0000-0002-4188-0572}}
\email[Email: ]{pr.mouhssine@gmail.com}
\affiliation{Quantum Physics and Magnetism Team, LPMC, Faculty of Science Ben
M'sik,\\
Casablanca Hassan II University,
Morocco.} 

\author{B. Mishra\orcidlink{0000-0001-5527-3565}}
\email[Email: ]{bivu@hyderabad.bits-pilani.ac.in}
\affiliation{Department of Mathematics, Birla Institute of Technology and
Science-Pilani,\\ Hyderabad Campus, Hyderabad-500078, India.}

\date{\today}

\begin{abstract}
In this paper, we have presented an accelerating cosmological model of the Universe in an extended symmetric teleparallel gravity or $f(Q,T)$ gravity. The parametric form of the Hubble parameter is, $H\left(z\right) =H_{0}\left[ \alpha +\left( 1-\alpha \right) \left( 1+z\right) ^{n}\right] ^{\frac{3}{2n}}$, where $H_0$ and $n$ are constants and for $n=3$, the $\Lambda$CDM scenario can be obtained. We have considered the logarithmic form of $f(Q,T)$ as, $f(Q,T)=-Q+\beta  \log\left(\frac{Q}{Q_{0}}\right)+\gamma  T$, where $\beta $ and $\gamma $ are the free model parameters. Using the Hubble, Baryon Acoustic Oscillations (BAO), and Type Ia Supernovae (SNe Ia) datasets, the present value of the Hubble parameter and other free parameters are constrained. Further other cosmographic and dynamical parameters are presented using the obtained constrained values of the Hubble and free parameters. The model shows the quintessence behavior of the Universe at the present time. The present value of the EoS parameter is obtained as, $\omega_{0}=-0.56$ for the $Hubble+BAO+SNe$ datasets. The energy conditions are presented and the violation of the strong energy condition has been shown. We have performed the dynamical system analysis to validate the stability of the model. From the evolutionary plot obtained through the dynamical system variables, the present value of density parameters have been obtained  as $\Omega_m\approx0.3$ and $\Omega_{de}\approx0.7$.
\end{abstract}

\maketitle

\section{Introduction}

\label{sec1}

The observations of Type Ia Supernovae (SNe Ia) \cite{SN1, SN2} in conjunction with Large Scale Structure (LSS) \cite{LSS1, LSS2}, Baryon Acoustic Oscillations (BAO) \cite{BAO1, BAO2}, and Cosmic Microwave Background (CMB) anisotropies \cite{CMB1, CMB2} have presented proof for cosmic acceleration. These findings resulted in a new kind of matter that violates the strong energy condition, namely, $\rho +3p<0$. Dark energy (DE) is an exotic matter that causes such a condition to be met at a given point in the evolutionary history of the Universe. This unexplained DE causes the late time cosmic acceleration with negative pressure. Observations also show that DE is the dominant component of the current Universe. According to the latest measurements from the Planck satellite mission, DE makes up approximately $73\%$ of the mass-energy budget of the Universe, while dark matter and ordinary baryonic matter make up $23\%$ and $4\%$, respectively \cite{Aghanim}. There are various possibilities that obeys the attribute of DE, such as quintessence \cite{Quin1, Quin2}, K-essence \cite{Ess1, Ess2}, Chaplygin gas models \cite{CG1, CG2}, and decaying vacuum models \cite{DV1, DV2, DV3, DV4}, and so on.

Modified gravity theories (MGT) have recently emerged as an alternative to standard cosmology and are becoming increasingly popular for describing the late-time cosmic acceleration mechanism. Geometrically, MGT are Einstein's General Theory of Relativity (GR) generalizations for which the Einstein-Hilbert action is changed by substituting the curvature scalar $R$ for a more generalized function. This may corresponds to a curvature scalar or a different function with matter-geometry coupling. Some extensively used alternative MGT are $f(R)$ gravity ($R$ be the curvature scalar) \cite{fR1, fR2, fR3}, $f(T)$ gravity ($T$ be the torsion scalar) \cite{fT1,fT2, fT3, fT4}, $f(Q)$ gravity ($Q$ be the non-metricity scalar) \cite{fQ1, fQ2,fQ3, fQ4, fQ5, fQ6, fQ7, fQ8}, and $f(Q,T)$ gravity ($T$ be the trace of stress energy-momentum tensor) \cite{fQT1}.

Among all these geometrically MGTs, the $f(Q,T)$ gravity has piqued the interest of several astrophysicists and cosmologists in recent years due to its ability to address a varieties of astrophysical and cosmological issues \cite{fQT1}. In this case, the gravitational
Lagrangian is expressed by an arbitrary function of the non-metricity scalar $Q$ and the trace of stress energy-momentum tensor $T$, the dependence of which can be induced by exotic imperfect fluids or quantum effects \cite{Harko}. The field equations were developed by varying the gravitational action with respect to both metric and connection. The matter-energy coupling in $f(Q,T)$ gravity has an important role in providing a thorough theoretical explanation of the late-time cosmic acceleration of the Universe without requiring the presence of DE. We discuss here, some of the prominent research that has shown 
that $f(Q,T)$ gravity can provide a promising framework to study the dynamics of the Universe on cosmological scales. The potential for $f(Q,T)$ gravity to provide a new perspective on the phenomenon of inflation in the early Universe has been shown in Ref. Shiravand et al. \cite{Inflation1} whereas this theory can naturally reproduce the observed power spectrum of cosmic microwave background radiation, while also addressing some of the shortcomings of the standard inflationary paradigm \cite{Inflation2}. The late time acceleration of the Universe can be realised without invoking the DE has been shown \cite{Koussour1,Koussour2}. The cosmological dynamics and the possibility of future singularity have been investigated in Ref. \cite{Pati,LP}, whereas the dynamical system analysis has been employed to investigate the evolution of density parameters \cite{LPati}. Nájera and Fajardo \cite{pert} have investigated cosmological perturbation theory in the context of $f(Q,T)$ gravity and revealed that the presence of non-minimal couplings between matter and curvature perturbations can significantly impact the evolution of the Universe. Some of the early Universe issues has been illustarted in the context of $f(Q,T)$ gravity in Ref. \cite{Agrawal, Najera}. One notable feature of $f(Q,T)$ gravity is that it violates the conservation of energy principle in its standard form. This is due to the non-minimal coupling between matter and curvature in the Lagrangian, which can result in the exchange of energy between these two components. However, it should be noted that various modifications have been proposed to the theory to address this issue, such as the introduction of a coupling term that can help to ensure the conservation of energy at the level of the field equations \cite{fQT1,Harko}. The Friedmann-Lema\^{\i}tre-Robertson-Walker (FLRW) metric is appropriate for representing the current state of the Universe since it is based on an isotropic and spatially homogeneous Universe. As a result of flat space-time, FLRW models are universally acceptable with the perfect fluid matter. In line with the literature, the FLRW cosmological model, perfect fluid matter, and various assumptions have been examined under $f(Q,T)$ gravity.

In this paper, we will investigate the logarithmic form of the function $f(Q,T)$ which has been studied in the background of other MGTs such as $f(R)$ gravity \cite{logR1, logR2, logR3}, $f(T)$ gravity \cite{logT1}, and $f(Q)$ gravity \cite{log}. It is very acceptable that when solving field equations or the so-called Friedmann equations in any theory of gravity, we generally assume that the Universe is filled with dust (i.e. $p=0$) and gets a solution. Otherwise, the solution is presupposed and confirmed by comparison with observational data. This last approach is known as the model-independent way of study of cosmological models or the cosmological parametrization \cite{Pacif1}. Generally, it presumes parametrizations of any kinematic variables such as the Hubble parameter $ H\left( t\right) $, deceleration parameter $q\left( t\right) $, jerk parameter $j\left( t\right) $ and EoS parameter $\omega \left( t\right)$ and so give an extra equation to completely solve the system of field equations Ref. \cite{H1, H2, H3, H4, q1, q2, q3, q4, j1, j2, j3, EoS1, EoS2, EoS3, EoS4}.

The paper is organized as follows: In Sec. \ref{sec2}, we have discussed the gravitational field equations of $f(Q,T)$ gravity for the flat FLRW Universe. In Sec. \ref{sec3}, we have introduced the $Hubble$, $BAO$, and $SNe~Ia$ datasets to constrain the free parameter of the parameterized Hubble parameter and analyse the behaviour of cosmographic parameters. In addition, we have also analysed the $Om(z)$ diagnostic. Further, in Sec. \ref{sec4}, we have employed the logarithmic form of $f(Q,T)$ and derived the dynamical parameters. Also we have examined the behavior of energy conditions incorporating the constrained values of the Hubble and free parameters. The dynamical stability analysis of the model has been analyzed in Sec. \ref{sec5} to obtain the stability of the model. Finally, in Sec. \ref{sec6}, we have given the results and conclusion.

\section{The field equations of $f(Q,T)$ Theory}

\label{sec2}

The $f(Q,T)$ gravity is a modified version of symmetric teleparallel gravity in which a matter Lagrangian $L_{m}$ may be characterized by the combination of $Q$ and $T$, where $Q$ and $T$ are scalars \cite{fQT1}. The total gravitational action of $f(Q,T)$ gravity is, 
\begin{equation}\label{1}
S=\int \sqrt{-g}\left( \frac{1}{2\kappa }f(Q,T)+L_{m}\right) d^{4}x,
\end{equation}
where $\kappa =8\pi$ and $g$ be the metric determinant. Without the loss of generality, We take $G=1$ and $c=1$. The energy-momentum tensor $T_{\mu \nu }$  can be written as,
\begin{equation}\label{2}
T_{\mu \nu }=-\frac{2}{\sqrt{-g}}\dfrac{\delta (\sqrt{-g}L_{m})}{\delta g^{\mu \nu }}.
\end{equation}
The variation of the energy-momentum tensor with respect to the metric tensor becomes, 
\begin{equation}\label{3}
\frac{\delta \,g^{\,\mu \nu }\,T_{\,\mu \nu }}{\delta \,g^{\,\alpha \,\beta }}=T_{\,\alpha \beta }+\Theta _{\,\alpha \,\beta }\,,
\end{equation}
and
\begin{equation}\label{4}
\Theta _{\mu \nu }=g^{\alpha \beta }\frac{\delta T_{\alpha \beta }}{\delta g^{\mu \nu }}.
\end{equation}
Further, the non-metricity scalar $Q$ expressed as \cite{fQ1, fQ2},
\begin{equation}\label{5}
Q\equiv -g^{\mu \nu }(L_{\,\,\,\alpha \mu }^{\beta }L_{\,\,\,\nu \beta}^{\alpha }-L_{\,\,\,\alpha \beta }^{\beta }L_{\,\,\,\mu \nu }^{\alpha }),
\end{equation}
where $L_{\,\,\,\alpha \gamma }^{\beta }$ denotes the disformation tensor, 
\begin{eqnarray}
L_{\alpha \gamma }^{\beta } &=&-\frac{1}{2}g^{\beta \eta }(\nabla _{\gamma}g_{\alpha \eta }+\nabla _{\alpha }g_{\eta \gamma }-\nabla _{\eta }g_{\alpha\gamma }).  \label{6} \\
&=&\frac{1}{2}g^{\beta \eta }\left( Q_{\gamma \alpha \eta }+Q_{\alpha \eta\gamma }-Q_{\eta \alpha \gamma }\right) ={L^{\beta }}_{\gamma \alpha }.\label{7}
\end{eqnarray}

The non-metricity tensor is written as, 
\begin{equation}\label{8}
Q_{\gamma \mu \nu }=-\nabla _{\gamma }g_{\mu \nu }=-\partial _{\gamma
}g_{\mu \nu }+g_{\nu \sigma }\widetilde{\Gamma }{^{\sigma }}_{\mu \gamma}+g_{\sigma \mu }\widetilde{\Gamma }{^{\sigma }}_{\nu \gamma },
\end{equation}
where $\widetilde{\Gamma }{^{\sigma }}_{\mu \gamma }$ is the Weyl--Cartan connection \cite{fQT1}, and the trace of the non-metricity tensor is presented as, 
\begin{equation}\label{9}
Q_{\beta }=g^{\mu \nu }Q_{\beta \mu \nu },\qquad \widetilde{Q}_{\beta
}=g^{\mu \nu }Q_{\mu \beta \nu }.
\end{equation}

A superpotential, also known as the non-metricity conjugate, can be described as, 
\begin{eqnarray}\label{10}
\hspace{-0.5cm} &&P_{\ \ \mu \nu }^{\beta }\equiv \frac{1}{4}\bigg[-Q_{\ \ \mu \nu }^{\beta }+2Q_{\left( \mu \ \ \ \nu \right) }^{\ \ \ \beta}+Q^{\beta }g_{\mu \nu }-\widetilde{Q}^{\beta }g_{\mu \nu }  \notag \\ 
\hspace{-0.5cm} 
&&-\delta _{\ \ (\mu }^{\beta }Q_{\nu )}\bigg]=-\frac{1}{2}%
L_{\ \ \mu \nu }^{\beta }+\frac{1}{4}\left( Q^{\beta }-\widetilde{Q}^{\beta}\right) g_{\mu \nu }-\frac{1}{4}\delta _{\ \ (\mu }^{\beta }Q_{\nu
)}.\quad\quad
\end{eqnarray}

Thus, the non-metricity scalar becomes \cite{fQ1}, 
\begin{eqnarray}\label{11}
&&Q=-Q_{\beta \mu \nu }P^{\beta \mu \nu }=-\frac{1}{4}\big(-Q^{\beta \nu \rho }Q_{\beta \nu \rho }+2Q^{\beta \nu \rho }Q_{\rho \beta \nu }  \notag \\
&&-2Q^{\rho }\tilde{Q}_{\rho }+Q^{\rho }Q_{\rho }\big).
\end{eqnarray}

By varying the action in Eqn. \eqref{1} with respect to the metric tensor components, one can get
\begin{multline}\label{12}
-\frac{2}{\sqrt{-g}}\nabla _{\beta }(f_{Q}\sqrt{-g}P_{\,\,\,\,\mu \nu
}^{\beta })-\frac{1}{2}fg_{\mu \nu }+f_{T}(T_{\mu \nu }+\Theta _{\mu \nu }) \\
-f_{Q}(P_{\mu \beta \alpha }Q_{\nu }^{\,\,\,\beta \alpha }-2Q_{\,\,\,\mu
}^{\beta \alpha }P_{\beta \alpha \nu })=8\pi T_{\mu \nu }.
\end{multline}
where $f_{Q}=\dfrac{df}{dQ}$ and $f_{T}=\dfrac{df}{dT}$.

In $f(Q,T)$ theory, the divergence of the matter-energy-momentum tensor can be expressed as:
\begin{eqnarray}\label{13}
\hspace{-0.5cm} &&\mathcal{D}_{\mu }T_{\ \ \nu }^{\mu }=\frac{1}{f_{T}-8\pi }
\Bigg[-\mathcal{D}_{\mu }\left( f_{T}\Theta _{\ \ \nu }^{\mu }\right) -\frac{16\pi }{\sqrt{-g}}\nabla _{\alpha }\nabla _{\mu }H_{\nu }^{\ \ \alpha \mu } 
\notag \\
\hspace{-0.5cm} &&+8\pi \nabla _{\mu }\bigg(\frac{1}{\sqrt{-g}}\nabla
_{\alpha }H_{\nu }^{\ \ \alpha \mu }\bigg)-2\nabla _{\mu }A_{\ \ \nu }^{\mu}+\frac{1}{2}f_{T}\partial _{\nu }T\Bigg],
\end{eqnarray}
where $H_{\gamma }^{\ \ \mu \nu }$ is the hyper-momentum tensor density
defined as,
\begin{equation}\label{14}
H_{\gamma }^{\ \ \mu \nu }\equiv \frac{\sqrt{-g}}{16\pi }f_{T}\frac{\delta T}{\delta \widetilde{\Gamma }_{\ \ \mu \nu }^{\gamma }}+\frac{\delta \sqrt{-g}\mathcal{L}_{M}}{\delta \widetilde{\Gamma }_{\ \ \mu \nu }^{\gamma }}.
\end{equation}

The matter-energy-momentum tensor in the $f(Q,T)$ gravity theory is not conserved, i.e. $\mathcal{D}_{\mu }T_{\ \ \nu }^{\mu }=B_{\nu} \neq 0$, as illustrated in Eqn. \eqref{13}. This non-conservation can be seen as an additional force that acts on massive test particles, leading to non-geodesic motion. It also indicates the amount of energy that enters or exits a specific volume of a physical system. Also, the non-zero right-hand side of the energy-momentum tensor implies the presence of transfer processes or particle production in the system. However, it should be noted that if $f_{T}$ terms are absent in the above equation, the energy-momentum tensor becomes conserved \cite{fQT1}.\\
We can obtain the the energy balance and the momentum conservation equations from the divergence of the energy-momentum tensor respectively as,
\begin{eqnarray}
    \dot{\rho}+3H(\rho+p) &=& B_{\mu}u^{\mu}~,\label{EB}\\
     \frac{d^{2}x^{\mu}}{ds^{2}}+\Gamma^{\mu}_{\alpha\beta}u^{\alpha}u^{\beta} &=& \frac{h^{\mu\nu}}{\rho+p}(B_{\nu}-\mathcal{D}_{\nu}p)~,\label{MC}
\end{eqnarray}
An over dot denotes the quantity as, $\dot{f} = u_{\mu}\mathcal{D}^{\mu}f$ and $H=\frac{1}{3}\mathcal{D}^{\mu}u_{\mu}$. The projection operator can be  denoted as, $h^{\mu\nu} = g^{\mu\nu}+u^{mu}u^{\nu}$. 
From the physical point of view, Eq. \eqref{EB} represents the energy balance of a gravitational system whereas Eq. \eqref{MC} be  the equation of motion of massive particles in terms of $f(Q,T)$. It represents the amount of energy entering or leaving a specified volume of a system. The source term $B_{\mu}u^{\mu}$ is for the energy creation or annihilation. If $B_{\mu}u^{\mu}=0$ satisfied  at all points in space time, then the total energy of a gravitating system is conserved. It is likely that when $B_{\mu}u^{\mu}\neq0$, the energy transfer processes or particle production occur in the given system.

Varying the gravitational action \eqref{1} with respect to the connection, we obtain the field equations as,
\begin{eqnarray} \label{15}
\nabla_{\mu}\nabla_{\nu}\bigg( \sqrt{-g}f_{Q} P^{\mu\nu}_{\ \ \ \ \beta}+4\pi H_{\beta}^{\ \ \mu \nu} \bigg)=0.
\end{eqnarray}

Now, to obtain the modified Friedmann equations, some constraints to be applied. Initially, we suppose that the matter content of the Universe is of perfect fluid. In addition, an energy-momentum tensor is a mathematical object that explains the distribution of matter and energy in space-time. The energy-momentum tensor for a perfect fluid is expressed as,
\begin{equation}\label{16}
T_{\mu \nu }=\left( \rho +p\right) u_{\mu }u_{\nu }+pg_{\mu \nu },
\end{equation}
where $\rho $, $p$, and $u^{\mu }$ represents respectively the energy density, pressure, and 4-vector velocity of the perfect fluid. Second, the FLRW metric is a mathematical representation of the geometry of the Universe. It represents a homogeneous and isotropic Universe, which means that the Universe seems the same at any point and in any direction in space-time. The metric is given as, 
\begin{equation}\label{17}
ds^{2}=-dt^{2}+a^{2}(t)\left[ dr^{2}+r^{2}d\theta ^{2}+r^{2}\sin ^{2}\left(\theta \right) d\varphi ^{2}\right]~,
\end{equation}
where $a(t)$ is the cosmological scale factor that represents the expansion of the Universe along the spatial directions in terms of the cosmic time $t$ with the current Universe $t=t_{0}$ as $a\left( t_{0}\right) =1$, $r$ denotes the comoving radial distance, $\theta $ and $\varphi $ denote the angular coordinates. With an adapted coordinate system, i.e. the spherical coordinates in the spatial variables of the line element \eqref{17} when the connection is zero (the coincident gauge), the non-metricity scalar is $Q=6H^{2}$, where $H$ is the Hubble parameter and also represent as a measure of the expansion of the Universe. Using the metric \eqref{17} and field equations \eqref{12}, the modified Friedmann equations are obtained as, 
\begin{eqnarray}
\kappa \rho &=& \frac{f}{2}-6FH^{2}-\frac{2\widetilde{G}}{1+\widetilde{G}}(\dot{F}H+F\dot{H}),\label{18}\\
\kappa p & =& -\frac{f}{2}+6FH^{2}+2(\dot{F}H+F\dot{H}),\label{19}
\end{eqnarray}
where the symbols $F=f_{Q}$, and $\kappa \widetilde{G}=f_{T}$, signify differentiation with respect to $Q$, and $T$, respectively.
The total density parameters can be given as,
\begin{equation}
\Omega_m+\Omega_r+\Omega_{de}=1
\end{equation}
where $\Omega_m$, $\Omega_r$ and $\Omega_{de}$ respectively denotes the density parameters for matter, radiation and dark energy phase. Here we consider $\Omega_r=0$, so that $\Omega_m+\Omega_{de}=1$.

It is worth noting that in the limit where $f(Q,T)$ reduces to the Einstein-Hilbert action and matter is described by a perfect fluid with negligible anisotropic stress i.e. $f(Q,T)=-Q$, the $f(Q,T)$ gravity equations reduce to the well-known Friedmann equations of the $\Lambda$CDM model.

\section{Observational Analysis}\label{sec3}
In the present study, we assume a two-parameter parametrization of the Hubble parameter $H(z)$ in terms of the redshift $z$, and attempt to derive all other important cosmological parameters, such as the EoS parameter $\omega $, the energy density and pressure. The following parametrization form of Hubble parameter is well-studied in literature \cite{Mamon} $H\left(z\right) =H_{0}\left[ \alpha +\left( 1-\alpha \right) \left( 1+z\right) ^{n}
\right] ^{\frac{3}{2n}}$, where $H_{0}$, $\alpha $, and $n$ are the current value of the Hubble parameter and arbitrary constants, respectively. It is important to note that the standard $\Lambda $CDM model conforms to the scenario for $n=3$, with the current cold dark matter density parameter $\Omega _{m}^{0}=\left( 1-\alpha \right) $. As a result, the model parameter $n$ is an excellent predictor of the deviation of the model from the $\Lambda$CDM model.

Using the above ansatz, the first derivative of the Hubble parameter with respect to cosmic time may be expressed in terms of redshift as, 
\begin{eqnarray*}
\overset{.}{H} &=&-\left( 1+z\right) H\left( z\right) \frac{dH\left(
z\right) }{dz}  \notag \\
&=&\frac{3}{2}H_{0}^{2}(\alpha -1)(z+1)^{n}\left[ \alpha -(\alpha
-1)(z+1)^{n}\right] ^{\frac{3}{n}-1}.
\end{eqnarray*}
Without DE, the Universe should decelerate because gravity binds stuff together. A cosmological model must include decelerated and accelerated expansion phases to reflect the complete evolutionary history of the Universe accurately. In this regard, the deceleration parameter is of utmost needed parameter and can be defined as,
\begin{equation}\label{20}
q=-1-\frac{\overset{.}{H}}{H^{2}}=-1+\frac{\left( 1+z\right) }{H\left(z\right) }\frac{dH\left( z\right) }{dz}.
\end{equation}

In addition, $q$ might be both positive and negative. Therefore, $q>0$ shows that the Universe is decelerating and that matter dominates DE. However, $q<0$ means that Universe is accelerating and the dominance is from the DE. Again, by using the previous ansatz and Eqn. \eqref{20}, we obtain
\begin{equation}\label{21}
q\left( z\right) =-1+\frac{3(1-\alpha )(1+z)^{n}}{2\left[ \alpha -(\alpha-1)(1+z)^{n}\right] }~.
\end{equation}
Also, the transition redshift $z_{tr}$ can be calculated as $q(z_{tr})=0$, 
\begin{equation}\label{22}
z_{tr} =-1+2^{1/n} \left(-\frac{\alpha }{\alpha -1}\right)^{1/n}~.
\end{equation}

In fact, for $n=3$ and by setting $\Omega _{m}^{0}=\left(1-\alpha \right) $, Eqn. \eqref{21} reduces to,
\begin{equation}\label{23}
q\left( z\right) =-1+\frac{3\Omega _{m}^{0}(1+z)^{3}}{2\left[ \left(
1-\Omega _{m}^{0}\right) +\Omega _{m}^{0}(1+z)^{3}\right] },
\end{equation}
which is equivalent to the $\Lambda $CDM model. So, our findings include the $\Lambda $CDM model which is good for characterizing the evolution of the Universe.

\subsection{Observational Constraints}
We discuss the cosmological constraints of the model under consideration. To overcome the problem of arbitrary values of model parameters, three types of observational datasets will be used such as, Hubble datasets \cite{Sharov}, BAO \cite{BAO1, BAO2, BAO3}, and SNe Ia \cite{Scolnic}. The statistical technique we employ allows us to constrain the parameters such as $ H_{0}$, $\alpha $, and $n$. We used the Markov Chain Monte Carlo (MCMC) method \cite{MCMC} in conjunction with the standard Bayesian methodology. The datasets listed below are used:

\subsubsection{Hubble}
As the first data available, we used an updated set of 57 Hubble data points. In this collection, 31 are measured using the differential age (DA) method, while the remaining 26 are measured using BAO and other methods in the redshift range $0.07\leq z\leq 2.42$ \cite{Sharov}. This allows us to determine the expansion rate of the Universe at the redshift $z$.
Hence, $H(z)$ can be estimated using $H\left( z\right) =-\frac{1}{1+z}\frac{dz}{dt}.$ Further, to obtain the mean values of the model parameters $H_{0}$, $\alpha $, and $n$, the chi-square ($\chi _{BAO}^{2}$) for Hubble datasets is written as, 
\begin{equation}\label{24}
\chi _{Hubble}^{2}=\sum_{j=1}^{57}\frac{\left[ H(z_{j})-H_{obs}(z_{j},H_{0},\alpha ,n)\right] ^{2}}{\sigma (z_{j})^{2}},
\end{equation}
where $H(z_{j})$ denotes the theoretical value of the Hubble parameter suggested by our cosmological model, $H_{obs}(z_{j},H_{0},\alpha ,n)$ the observed value, and $\sigma (z_{j})$ the standard error in the measured value of $H\left( z\right) $.

\subsubsection{BAO}
In the case of BAO data available, we present results from the SDSS, 6dFGS, BOSS-DR12, and Wiggle Z surveys at different redshifts. To obtain BAO constraints, we use these expressions for measurable quantities: \newline
\begin{eqnarray}
d_{A}(z) &=& c\int_{0}^{z}\frac{d\widehat{z}}{H(\widehat{z},H_{0},\alpha ,n)},\label{25}\\
D_{v}(z) &=& \left[ \frac{d_{A}^{2}\left( z\right) cz}{H(z)}\right] ^{\frac{1}{3}},\label{26}
\end{eqnarray}
where $d_{A}(z)$ is the distance of comoving angular diameter and $D_{v}(z)$ is the dilation scale. The chi-square ($\chi _{BAO}^{2}$) for BAO datasets is written as,\newline
\begin{equation}\label{27}
\chi _{BAO}^{2}=Y^{T}C_{BAO}^{-1}Y.
\end{equation}
\newline
Here, $Y$ depends on the survey considered and $C_{BAO}$ is the covariance matrix (please see \cite{BBAO1, BBAO2, BBAO3}).\newline

\subsubsection{SNe Ia}
The observation of SNe Ia is critical for
understanding how the Universe is expanding. The Panoramic Survey Telescope and Rapid Response System (Pan-STARSS1), Sloan Digital Sky Survey (SDSS), Supernova Legacy Survey (SNLS), and Hubble Space Telescope (HST) surveys all collected data on SNe Ia \cite{Scolnic}. We employ the Pantheon sample, which consists of 1048 points with distance modulus $\mu _{obs}$ in the range $0.01<z_{j}<2.26$ at various redshifts. The distance modulus can be
calculated using the following equations:%
\begin{equation}\label{28}
\mu _{th}(z_{j})=25+5log_{10}\left[ \frac{d_{l}\left( z\right) }{1Mpc}\right],
\end{equation}
where
\begin{equation}\label{29}
d_{l}(z)=c(1+z)\int_{0}^{z}\frac{d\widehat{z}}{H(\widehat{z},H_{0},\alpha ,n)}.
\end{equation}

The chi-square ($\chi _{SNeIa}^{2}$) function, which connects model
predictions to SNe Ia data is written as,\newline
\begin{equation}\label{30}
\chi _{SNeIa}^{2}=\sum_{j,i=1}^{1048}\Delta \mu
_{j}(C_{SNeIa}^{-1})_{ji}\Delta \mu _{i}~,
\end{equation}
where, $\Delta \mu _{j}=\mu _{th}(z_{j},H_{0},\alpha ,n)-\mu _{obs}$, and $ C_{SNeIa}$ is the covariance matrix (please see \cite{Scolnic}).

\subsubsection{Results}
The analysis presented here provides important constraints on the model parameters for the combined $Hubble+BAO+SNe$ datasets. The combined likelihood function ($\mathcal{L}\propto e^{-\frac{\chi ^{2}}{2}}$): $\mathcal{L}%
_{Hubble}^{2}+\mathcal{L}_{BAO}^{2}+\mathcal{L}_{SNe}^{2}$, which takes into account the constraints from Hubble parameter measurements, BAO, and SNe Ia observations, has been used to derive the corresponding $\chi ^{2}$ values: $\chi _{Hubble}^{2}+\chi _{BAO}^{2}+\chi _{SNe}^{2}$. The results are displayed in TABLE -- \ref{tab} and FIG. \ref{H+SN+BAO}, which show the best-fit values and the confidence intervals for the parameters. The MCMC technique from the python package \textit{emcee} \cite{MCMC} has been used for likelihood minimization, which is a widely used tool in cosmology to explore the parameter space. The likelihood contours in FIG. \ref{H+SN+BAO} demonstrate the correlation between the different parameters and provide insights into the degeneracies in the model. Furthermore, the error bars for the Hubble parameter measurements and the distance modulus are displayed in FIG. \ref{ErrorHubble} and FIG. \ref {ErrorSNe}, respectively. These error bars reflect the uncertainties in the observations and are crucial for assessing the accuracy of the model predictions. The present analysis provides strong support for the current understanding of the universe's expansion history and the underlying cosmological model. The present Hubble parameter value $\left( z=0\right) $ is evaluated to be $H_{0}=65.39_{-0.83}^{+0.82}$ for the $Hubble+BAO+SNe$ datasets, which is in agreement with recent Planck findings \cite{Aghanim} and other studies in a similar context \cite{Chen1, Chen2, Aubourg, Capozziello}. These results contribute to our understanding of the evolution of Universe and provide crucial insights into the nature of DE, which remains some of the most pressing questions in modern cosmology.

\begin{widetext}

\begin{figure}[h]
\centerline{\includegraphics[scale=0.75]{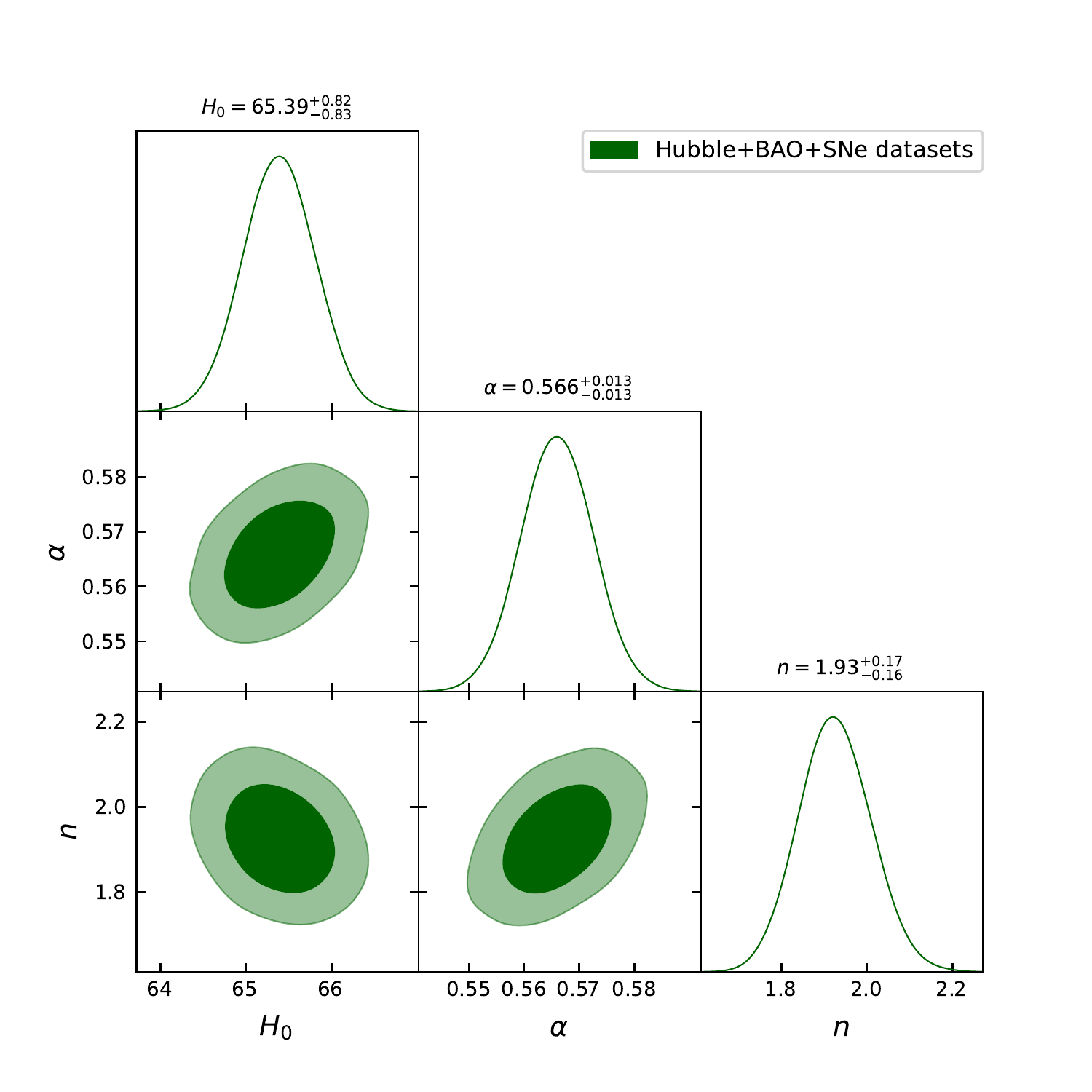}}
\caption{The figure displays the confidence ranges of parameter space for the $Hubble+BAO+SNe$ datasets, with $1-\sigma$ and $2-\sigma$ confidence intervals highlighted.}
\label{H+SN+BAO}
\end{figure}

\begin{figure}[H]
\centerline{\includegraphics[scale=0.60]{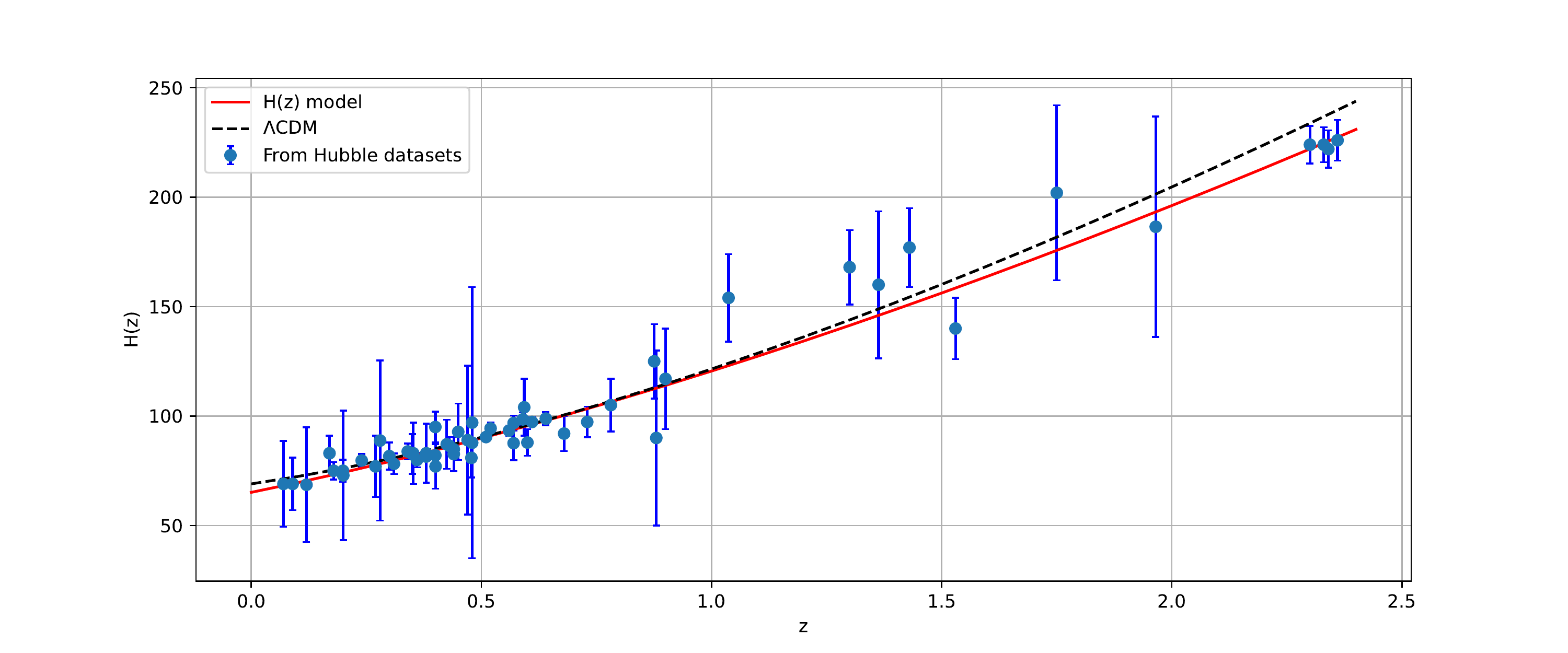}}
\caption{
The figure shows the Hubble parameter $H(z)$ as a function of redshift $z$ for the Hubble datasets (red line), along with the expected Hubble parameter values for a $\Lambda$CDM model (black dashed line). The error bar represents the uncertainties in the 57 data points used to construct the Hubble dataset.}
\label{ErrorHubble}
\end{figure}

\begin{figure}[h]
\centerline{\includegraphics[scale=0.60]{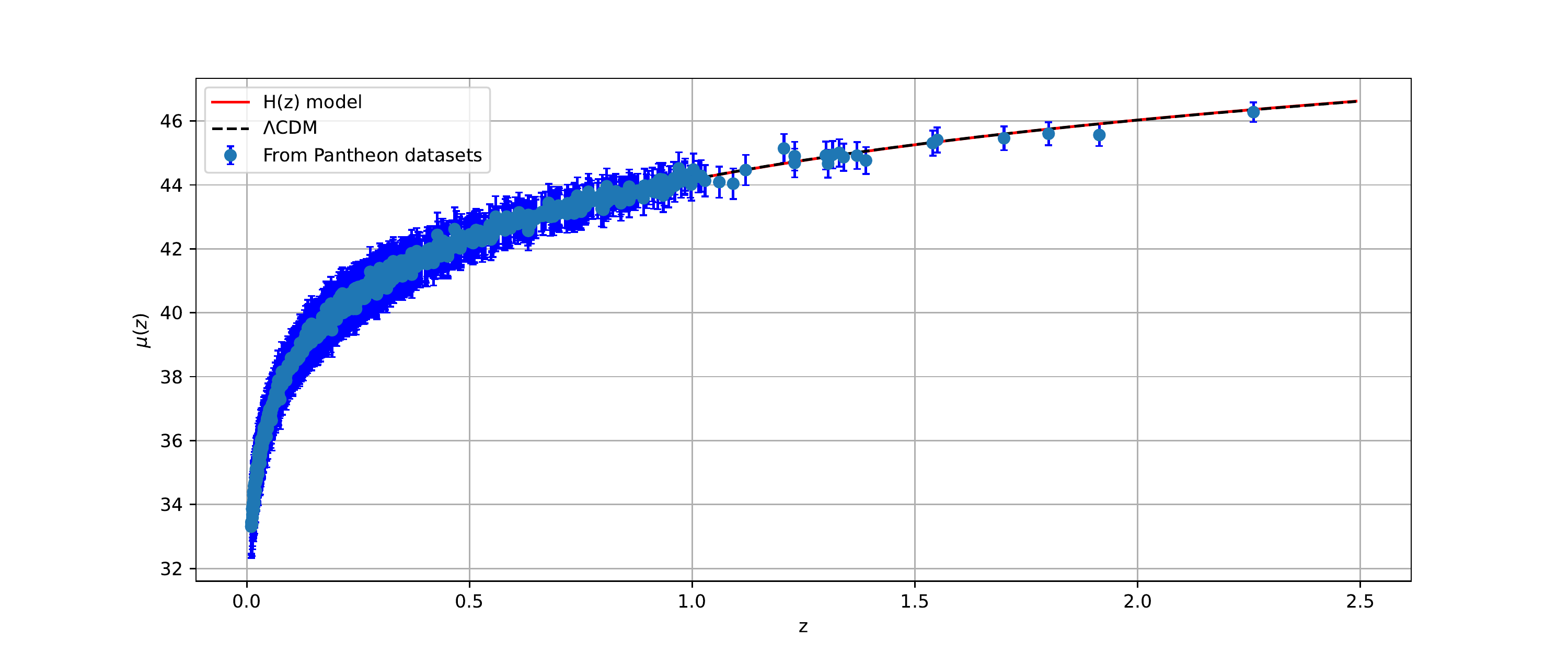}}
\caption{
The figure shows the distance modulus $\mu(z)$ as a function of redshift $z$ for the Pantheon datasets (red line), along with the expected $\mu(z)$ values for a $\Lambda$CDM model (black dashed line). The error bar represents the uncertainties in the 1048 data points used to construct the Pantheon dataset.}
\label{ErrorSNe}
\end{figure}

\begin{table*}[!htb]
\begin{center}
\begin{tabular}{l c c c c c c c c c}
\hline 
$Datasets$  & $H_{0}$ & $\alpha$ & $n$ & $q_{0}$ & $z_{tr}$ & $\omega_{0}$\\
\hline
$Priors$ & $(60,80))$  & $(0,10)$  & $(0,10)$ & $-$ & $-$ & $-$\\
$H+SNe+BAO$   & $65.39^{+0.82}_{-0.83}$  & $0.566^{+0.013}_{-0.013}$  & $1.93^{+0.17}_{-0.16}$ & $-0.35^{+0.02}_{-0.02}$ & $0.64^{+0.02}_{-0.02}$ & $-0.56^{+0.39}_{-0.39}$\\

\hline
\end{tabular}
\caption{The best-fit ranges of parameter space for the $Hubble+BAO+SNe$ datasets.}
\label{tab}
\end{center}
\end{table*}
\end{widetext}

\subsection{Deceleration parameter}

\begin{figure}[h]
\centerline{\includegraphics[scale=0.7]{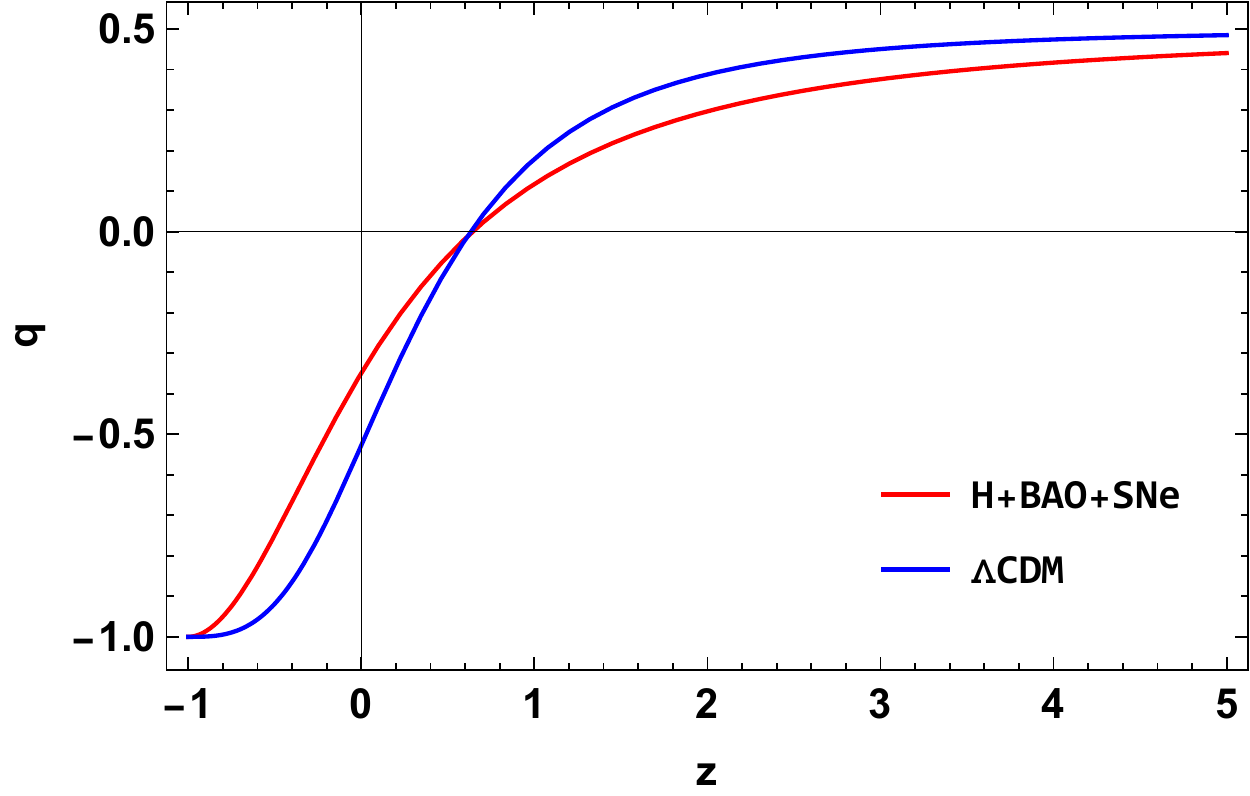}}
\caption{Plot of deceleration parameter $q$ versus redshift $z$ using the combined $Hubble+BAO+SNe$ datasets.}
\label{F_q}
\end{figure}

In this subsection, we will investigate the evolution of significant cosmological parameters and compare our calculated values with the observable evidence. Further, we have used the MCMC approach to constrain the parameters $H_{0}$, $\alpha $, and $n$ using the $Hubble+BAO+SNe$ datasets, as shown in TABLE -- \ref{tab}. In FIG. \ref{H+SN+BAO}, we show the best-fit values for the parameters $H_{0}$, $ \alpha $ and $n$. Because the model parameter $\gamma $ is not explicitly contained in the Hubble parameter, we fix it in order to examine the history of density, pressure, and EoS parameter. So, we chose the value  $\gamma =0.2$, which is constrained by observational datasets.  The deceleration parameter ($q$) has three parameters $H_{0}$, $\alpha $, and $n$ that are constrained by the combined $Hubble+BAO+SNe$ datasets, thus we can now describe its evolution using numerical values. FIG. \ref{F_q} depicts the evolution of $q$ with respect to redshift $z$, explaining its evolution in the recent past, current evolution, and signature flipping behavior given the aforementioned constrained values of the model parameters $H_{0}$, $\alpha $ and $n$. While the negative value of $q$ represents the accelerated phase, the positive value of $q$ represents the decelerating phase. The deceleration parameter $q
$, as shown in FIG. \ref{F_q}, changes with $z$ from positive to negative. This indicates a transition behaviour of early deceleration to late time acceleration at redshift $z_{tr}=0.64^{+0.02}_{-0.02}$ according to the constrained values of model parameters by the $Hubble+BAO+SNe$ datasets. It should be noted that the current value of $q$ for $Hubble+BAO+SNe$ datasets is $q_{0}=-0.35^{+0.02}_{-0.02}$ \cite{Capozziello}.

\subsection{State finder analysis}
State finder analysis is a cosmological approach to study the dynamics of the Universe. It is an important approach to understand different DE models. The state finder parameter are specified as a mixture of the Hubble parameter and its temporal derivatives. Sahni et al. \cite{Sahni1} proposed the state finder analysis as an extension of the EoS parameter ($\omega $). The state finder parameters $r$ and $s$ are as follows:%
\begin{eqnarray}
r &=& \frac{\dddot{a}}{aH^{3}}=2q^{2}+q-\frac{\dot{q}}{H}~,\label{31}\\
s &=& \frac{(r-1)}{3(q-\frac{1}{2})}~.\label{32}
\end{eqnarray}

The state finder parameter $r$ quantifies the departure of cosmic expansion from a pure power-law behavior, whereas the parameter $s$ defines its acceleration. The state finder analysis out-performs other cosmological parameters such as the EoS parameter and the distance-redshift relation. It is a geometrical diagnostic that makes no assumptions about the underlying cosmological hypothesis. Second, it is a model-independent approach for distinguishing between various DE scenarios, such as the $\Lambda $CDM (Lambda Cold Dark Matter), SCDM (Standard Cold Dark Matter), HDE (Holographic Dark Energy), CG (Chaplygin Gas) and quintessence. Lastly, it can act as a check on the assumptions made by other cosmological parameters. Here are some DE scenarios with corresponding values of $r$ and $s$ parameters,
\begin{itemize}
\item ${r=1,s=0}$ denotes $\Lambda $CDM scenario,

\item ${r=1,s=1}$ denotes SCDM scenario,

\item ${r=1,s=}\frac{2}{3}$ denotes HDE scenario,

\item ${r>1,s<0}$ denotes CG scenario,

\item ${r<1,s>0}$ denotes Quintessence scenario.\newline
\end{itemize}

For our model, the state finder parameters are expressed as,
\begin{eqnarray}\label{33}
r\left( z\right) &=
\frac{2\alpha ^{2}-\left[ (\alpha -1)\alpha (3n-5)(1+z)^{n}%
\right] +2(\alpha -1)^{2}(1+z)^{2n}}{2\left( \alpha -(\alpha
-1)(1+z)^{n}\right) ^{2}},
\end{eqnarray}
\begin{eqnarray}\label{34}
s\left( z\right) &= -\frac{(\alpha -1)(n-3)(1+z)^{n}}{3(\alpha
-1)(1+z)^{n}-3\alpha }.
\end{eqnarray}

\begin{figure}[h]
\centerline{\includegraphics[scale=0.70]{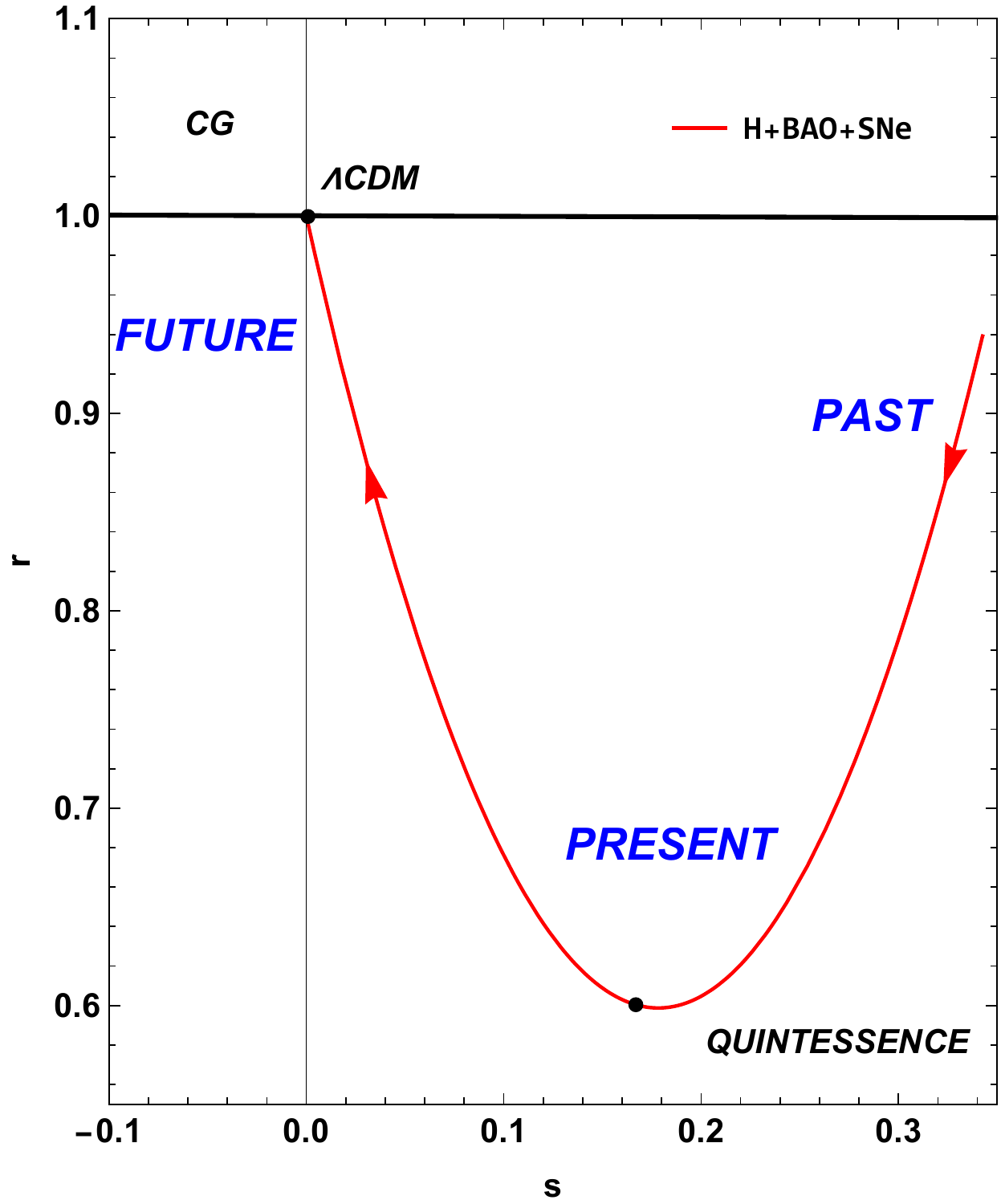}}
\caption{Plot of $r-s$ plane using the combined $Hubble+BAO+SNe$ datasets with $-1\leq z\leq5$.}
\label{F_rs}
\end{figure}

\begin{figure}[h]
\centerline{\includegraphics[scale=0.70]{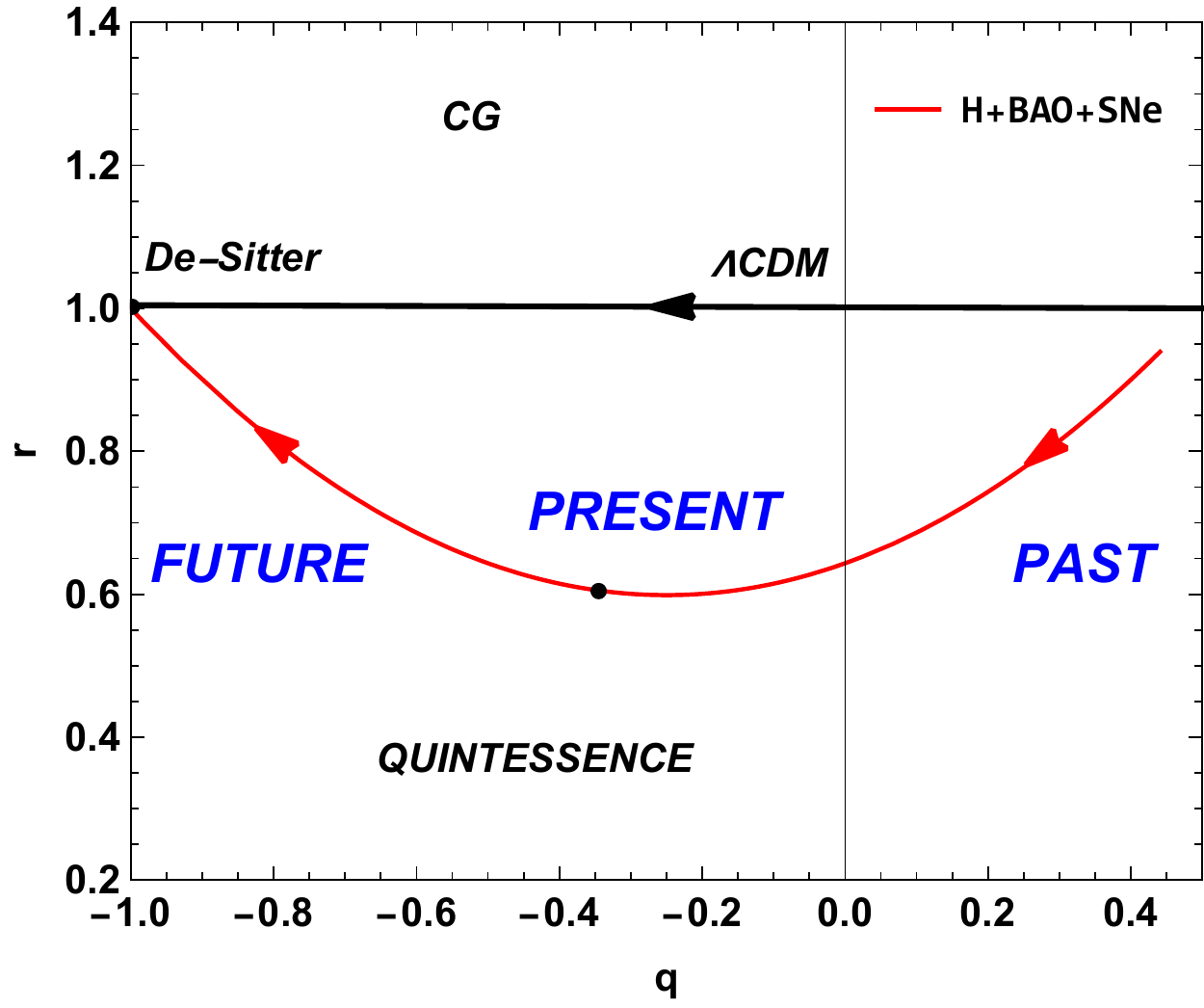}}
\caption{Plot of $r-q$ plane using the combined $Hubble+BAO+SNe$ datasets with $-1\leq z\leq5$.}
\label{F_rq}
\end{figure}

FIG. \ref{F_rs} shows the evolution of the model in the $r-s$ plane for the combined datasets. Our model follows the quintessence scenario since the trajectory is in the area $r<1$ and $s>0$. The trajectory of our model approaches the $\Lambda$CDM fixed point in the future, indicating that the Universe will continue to expand at an accelerating rate. This behavior is consistent with current observations and provides support for the idea that DE, represented by the cosmological constant $\Lambda$, is the driving force behind this acceleration. In addition, from the $r-q$ graph (FIG. \ref{F_rq}) it is clear that the deceleration parameter will continue to decrease with time, eventually approaching a value of $q = -1$, indicating a transition from acceleration to a period of constant expansion. This transition, known as the De-Sitter phase ($r=1$ and $q=-1$), is a crucial feature of modern cosmological models and is supported by a range of observational evidence.

\subsection{$Om(z)$ diagnostic}
In addition to the state finder analysis, a new diagnostic named $Om(z)$ has recently been presented \cite{Sahni2}, which aids in distinguishing the current matter density contrast $\Omega _{m}^{0}$ in various models more successfully. $Om(z)$, a novel DE diagnostic, is developed to distinguish $\Lambda$CDM from other DE scenarios. The $Om(z)$ diagnostic begins with the Hubble parameter, which is specified in \cite{Sahni2}, 
\begin{equation}\label{35}
Om\left( x\right) =\frac{h^{2}\left( x\right) -1}{x^{3}-1},
\end{equation}
in which $x=1+z$ and $h\left( x\right) =\frac{H\left( x\right) }{H_{0}}$. The $Om(z)$ diagnostic, which only includes the first derivative of the scale factor through the Hubble parameter, is thus simple to reconstruct from observational data. $Om(z)=\Omega _{m}^{0}$ is a constant in the $\Lambda $CDM model, independent of redshift $z$. It gives a cosmological constant null test. The advantage of the $Om(z)$ diagnostic is that it can differentiate DE models with less dependence on matter density $\Omega _{m}^{0}$ than the EoS
of DE \cite{Sahni2}. Ref. \cite{om} also discusses $Om(z)$ and state finder
diagnostics for the generalized CG model and the decaying vacuum model based on cosmic data. So, here we obtain
\begin{equation}\label{36}
Om\left( z\right) =\frac{\left[ \alpha -(\alpha -1)(1+z)^{n}\right] ^{3/n}-1}{(1+z)^{3}-1}.
\end{equation}

\begin{figure}[h]
\centerline{\includegraphics[scale=0.70]{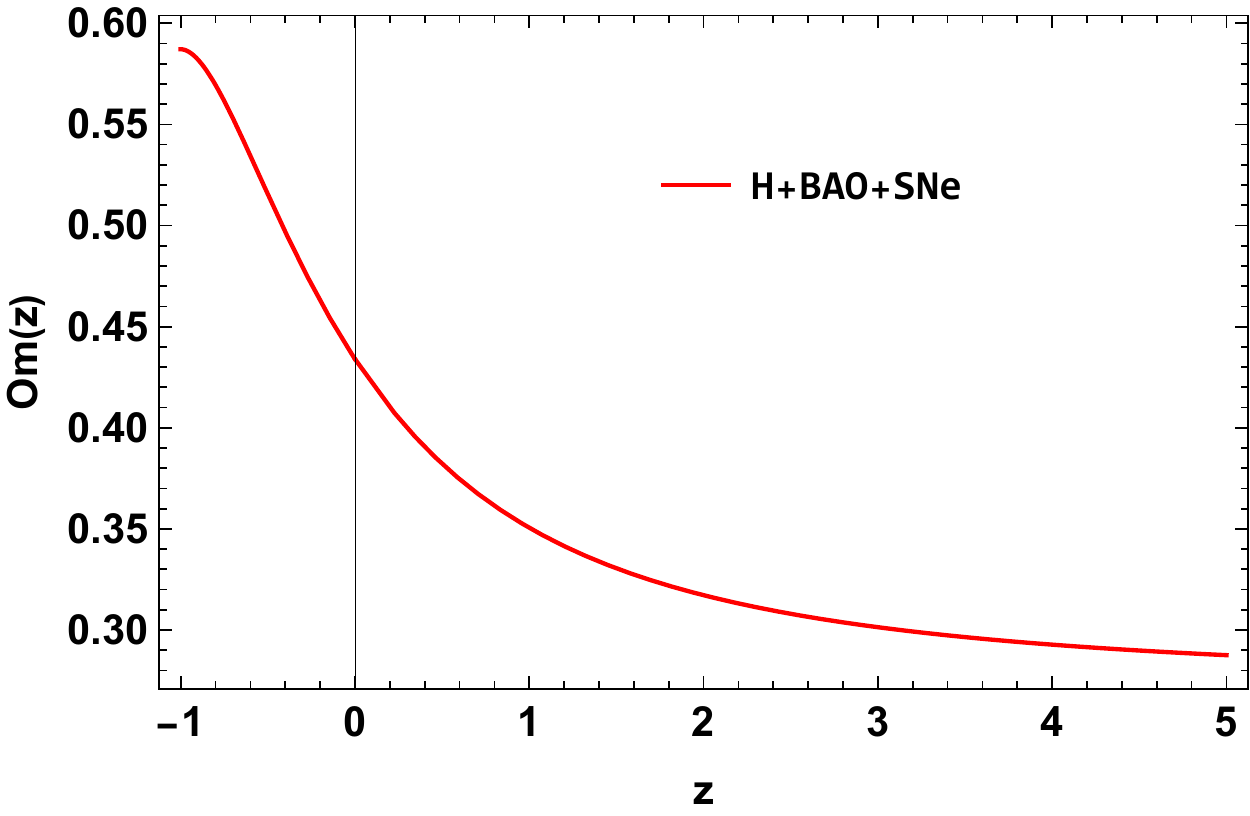}}
\caption{Plot of $Om(z)$ versus redshift $z$ using the combined $Hubble+BAO+SNe$ datasets.}
\label{F_Om}
\end{figure}

FIG. \ref{F_Om} shows the evolution of $Om(z)$ versus redshift $z$ for the $Hubble+BAO+SNe$ datasets. It can be shown that $Om(z)$ decreases as $z$ increases, implying that $Om(z)$ increases owing to Universe expansion in our model. So according to Ref. \cite{Sahni2}, a positive slope of $Om(z)$ indicates phantom ($\omega <-1$) while a negative slope indicates quintessence ($\omega >-1$). FIG. \ref{F_Om} shows that the $%
Om(z)$ has a negative slope, demonstrating quintessence-like behavior in our model.

\section{Logarithmic $f(Q,T)$ Cosmology}\label{sec4}
Typically it is preferable to have the exact solutions of the field equations to construct the cosmological model of the Universe. Unfortunately, the very non-linear equivalent Friedmann equations make the process quite challenging and hence the same behaviour in the corresponding modified Friedmann equations of the extended symmetric teleparallel gravity model. Therefore,  we consider some feasible assumptions to solve the system. The logarithmic model have shown some significant results in teleparallel gravity \cite{logmodel}. Motivated from this, we consider a particular model as $f(Q,T)=-Q+\beta  \log\left(\frac{Q}{Q_{0}}\right)+\gamma  T$, where $\beta$ and $\gamma$ are the free model parameters such that, $F=f_{Q}=-1+\frac{\beta}{Q}$, $\kappa \widetilde{G}=\gamma$ and $Q_{0}=6H_{0}^{2}$. It is essential to note that $\beta=0$ and $\gamma=0$ correspond to $f(Q,T)=-Q$, which is the case of GR theory. Moreover, for $T=0$, the theory is reduces to $f(Q)=-Q+\beta  \log \left(\frac{Q}{Q_{0}}\right)$ gravity. In this work, we have extended the investigation of the logarithmic model in symmetric teleparallelism with the additional term of the trace of an energy-momentum tensor. The free parameters of the Hubble parameter were constrained using the observational datasets as in Ref. \cite{log}. Here we aim to provide a more comprehensive analysis of the behaviour of the model by exploring its implications for the underlying cosmology in the context of $f(Q,T)$ gravity. In the present setting, the energy density and pressure expressions for this can be derived using Eqn. \eqref{18} and  Eqn. \eqref{19} as,

\begin{widetext}
\begin{eqnarray}
\rho &=& \frac{1}{\phi H^{2}}\left[3H^{2}(\gamma+8\pi)\left(-2\beta+\beta\log\left(\frac{H^{2}}{H_{0}^{2}}\right)+6H^{2}\right)-\gamma\left(\beta+6H^{2}\right)\dot{H}\right], \label{37}\\
p &=& -\frac{1}{\phi H^{2}}\left[3H^{2}\gamma+8\pi)\left(-2\beta+\beta\log\left(\frac{H^{2}}{H_{0}^{2}}\right)+6H^{2}\right)+(3\gamma+16\pi)\left(\beta+6H^{2}\right)\dot{H}\right]~.\label{38}
\end{eqnarray}

Using \eqref{37} and \eqref{38}, we obtain the EoS parameter 
\begin{equation}\label{39}
\omega =\frac{p}{\rho }= -1+\frac{4\dot{H}(\gamma +4\pi )(6H^{2}+\beta)}{\beta\gamma \dot{H}-18H^{4}(8\pi+\gamma)+6H^{2}(8\pi\beta+\dot{H}+\beta)\gamma)-3H^{2}\beta(8\pi+\gamma)\log\left(\frac{H^{2}}{H_{0}^{2}}\right)},
\end{equation}
\end{widetext}

where $\phi =12(\gamma +4\pi )(\gamma +8\pi )$.

\begin{figure}[h]
\centerline{\includegraphics[scale=0.50]{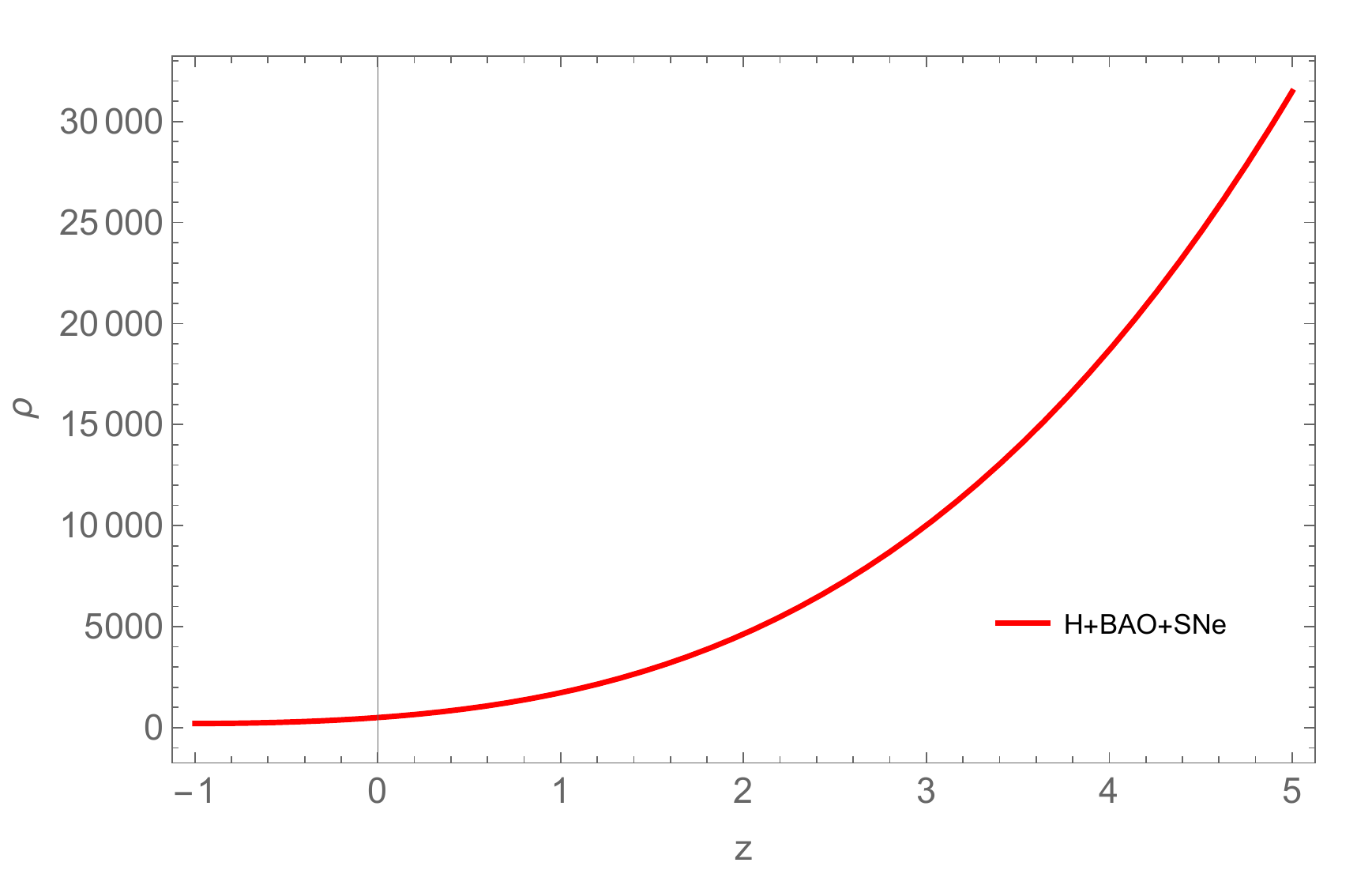}}
\caption{Plot of energy density $\protect\rho$ versus redshift $z$ using the combined $Hubble+BAO+SNe$ datasets.}
\label{F_rho}
\end{figure}

\begin{figure}[h]
\centerline{\includegraphics[scale=0.50]{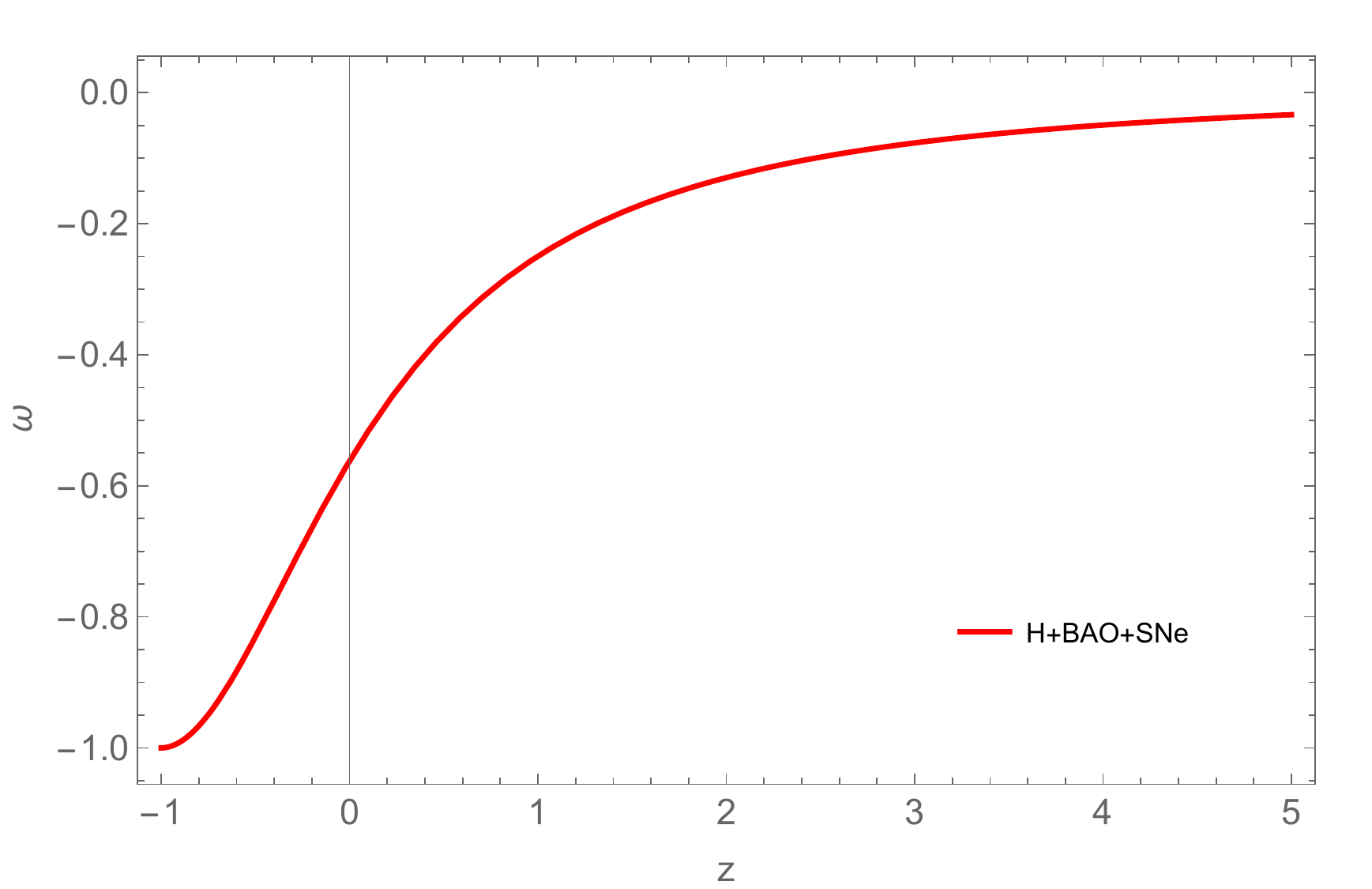}}
\caption{Plot of EoS parameter $\protect\omega$ versus redshift $z$ using the combined $Hubble+BAO+SNe$ datasets.}
\label{F_EoS}
\end{figure}

According to FIG. \ref{F_rho}, which depicts the evolution of the energy density, we observe that the energy density remains positive throughout the evolution and increases as the redshift $z$ increases. At the beginning of the evolution, the energy density has a positive value and decreases as $z$ approaches $-1$. This behavior is consistent with the expected behavior of the energy density of DE, which is expected to be non-zero and positive at all times and plays a crucial role in driving the accelerating expansion of the Universe. The EoS parameter $\omega $ is defined as the ratio of pressure to fluid energy density i.e. $p=\omega \rho $ and is not always constant. The vacuum energy with $\omega _{\Lambda }=-1$ is the simplest DE possibility and is comparable to the cosmological constant ($\Lambda $). Alternatives to vacuum energy are quintessence ($\omega >-1$), phantom energy ($\omega <-1$), and quintom (that can pass from phantom area to quintessence area as developed) and have time-dependent EoS parameter. The EoS parameter shown in FIG. \ref {F_EoS} suggests that the cosmic fluid behaves as quintessence DE. Moreover, the current value of the EoS parameter for the $Hubble+BAO+SNe$ datasets is $\omega_{0}=-0.56^{+0.39}_{-0.39}$.
\begin{figure}[h]
\centerline{\includegraphics[scale=0.50]{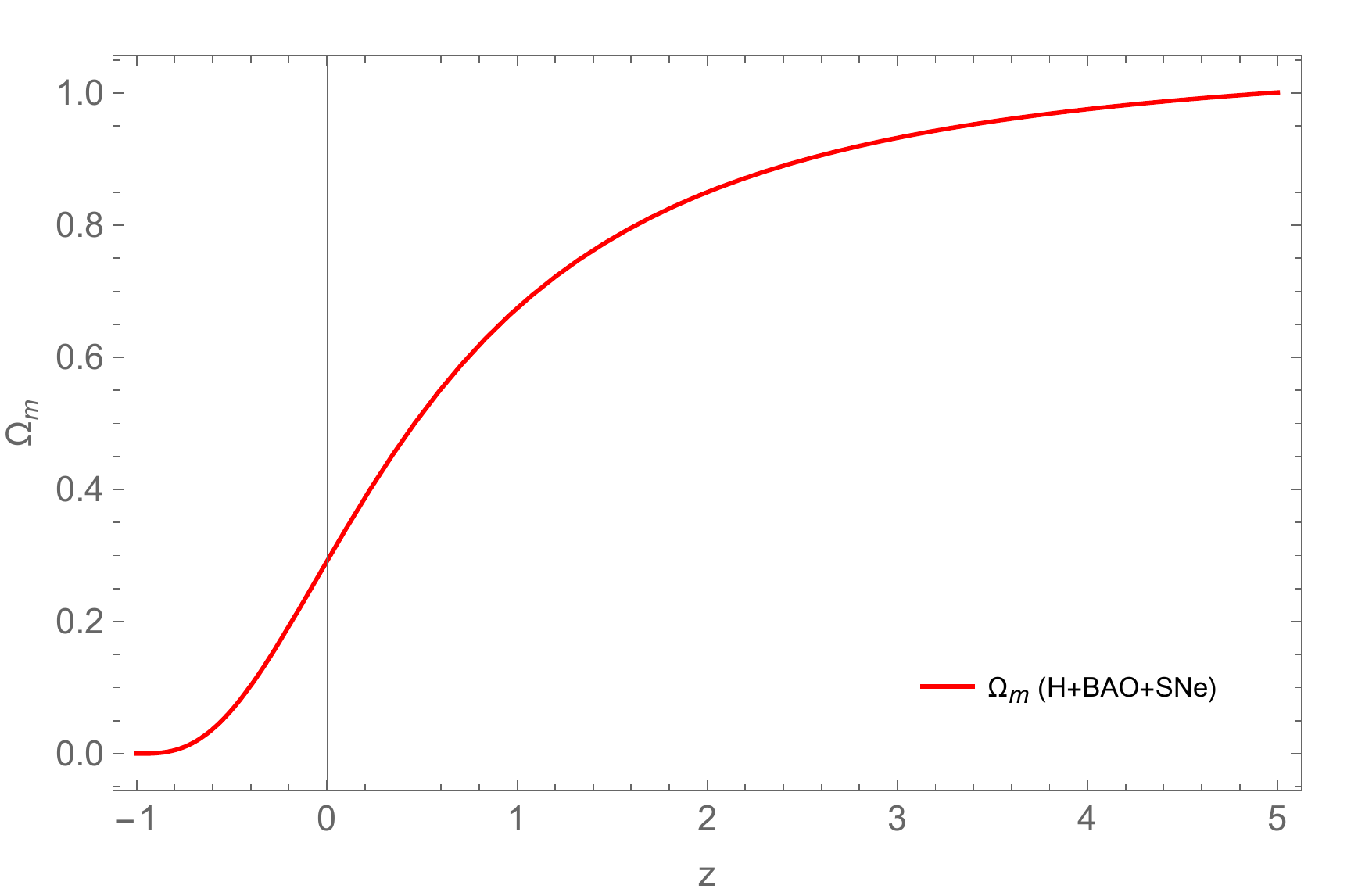}}
\caption{Plot of density parameter $\Omega$ versus redshift $z$ using the combined $Hubble+BAO+SNe$ datasets.}
\label{omegam0}
\end{figure}
The density parameter is given by $\Omega(z) = \frac{8\pi G\rho}{3H^{2}}$. The present density for matter is given by $\rho_{m}=\rho_{m0}(1+z)^{3}$. Therefore using Eqn. \eqref{37}, we get
\begin{eqnarray*}
    \Omega_{m}(z)=\frac{8\pi G\rho_{m0}(1+z)^{3}}{3H^{2}}~.
\end{eqnarray*}
Using the best-fit values obtained in TABLE -- \ref{tab}, the behavior of $\Omega_{m}$ is given in FIG. \ref{omegam0}. We can see that at present $\Omega_{m}$ gives the value $\Omega_{m0}\approx 0.291132$ which is very close to the observational values.

\subsection{Energy conditions}
In this subsection, we shall analyse the behavior of energy conditions (ECs) with the constrained free parameters in the context of $f(Q,T)$ gravity model. The motivation is to check the violation of strong energy condition \cite{EC}, which has been inevitable in the accelerating behaviour of the Universe in modified gravity. We shall derive the expressions for the energy conditions in redshift, and their violation or agreement at different epochs. In GR, the ECs are a set of physical constraints that connects the distribution of matter and energy to the curvature of space-time. These conditions are essential to comprehending the behavior of matter and energy in the context of modified gravity. The ECs impose a unique set of constraints on the distribution of matter and energy in space-time. For a perfect fluid type
stress-energy tensor $T_{\mu \nu }=(p+\rho )u_{\mu }u_{\nu }+pg_{\mu \nu }$. The ECs are \cite{ECs1, ECs2, ECs3}:

\begin{itemize}
\item WEC: Weak energy condition, $\rho \geq 0$, $\rho +p\geq 0$;

\item NEC: Null energy condition, $\rho +p\geq 0$;

\item DEC: Dominant energy condition, $\rho \pm p\geq 0$;

\item SEC: Strong energy condition, $\rho+3p\ge 0$, $\rho+p\ge0$.
\end{itemize}

One can see that the violation of NEC leads to the violation of residual ECs, which indicates the decrease of energy density as the Universe expands. Moreover, the violation of the SEC indicates the acceleration of the Universe. Now, using Eqns. \eqref{37} and \eqref{38}, we derive the expressions of ECS as,

\begin{widetext}

\begin{eqnarray}
\rho +p &=& -\frac{\overset{.}{H}(6H^{2}+\beta)}{3H^{2}(\gamma+8\pi )}\geq 0~,\label{40}\\
\rho-p &=& \frac{18H^{4}-6H^{2}\beta+\dot{H}(6H^{2}+\beta)+3H^{2}\beta\log\left(\frac{H^{2}}{H_{0}^{2}}\right)}{6H^{2}(\gamma+4\pi )}\geq 0~,\label{41}\\
\rho +3p &=& -\frac{6H^{2}(3H^{2}-\beta)(8\pi+\gamma)+\dot{H}(6H^{2}+\beta)(24\pi+5\gamma)+3H^{2}\beta(8\pi+\gamma)\log\left(\frac{H^{2}}{H_{0}^{2}}\right)}{6H^{2}(\gamma+4\pi )(8\pi+\gamma)}\geq 0~.\label{42}
\end{eqnarray}
\end{widetext}

\begin{figure}[h]
\centerline{\includegraphics[scale=0.50]{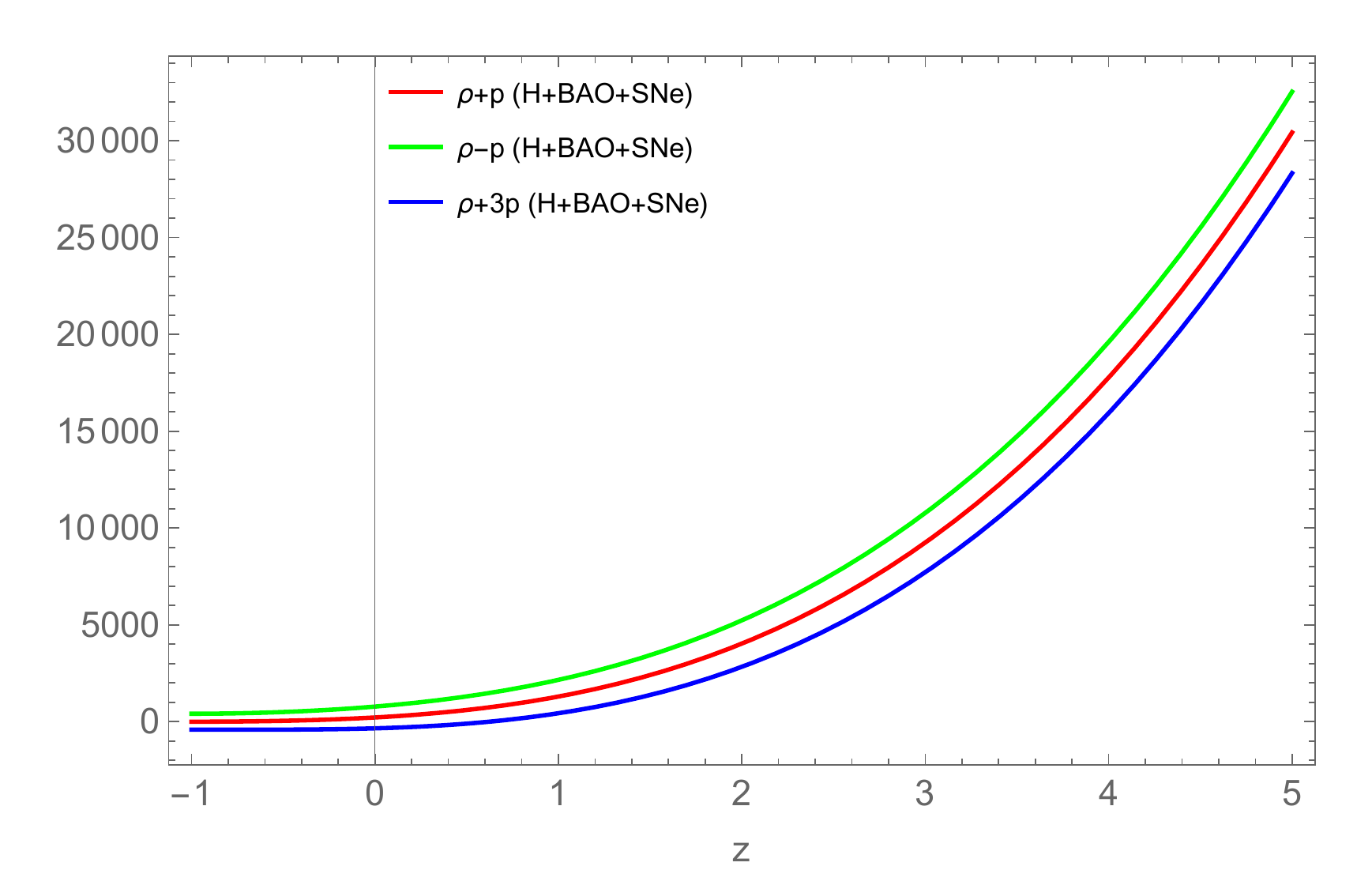}}
\caption{Plot of ECs versus redshift $z$ using the combined $Hubble+BAO+SNe$ datasets.}
\label{F_ECs}
\end{figure}
ECs for the logarithmic $f(Q,T)$ model are presented in FIG. \ref{F_ECs}. Furthermore, we can see from the figure that $\rho +p\geq 0$ and $\rho -p\geq 0$, which indicate that the NEC and DEC are satisfied. Additionally, $\rho +3p\leq 0$ leads to a violation of the SEC. As a result, the violation of the SEC causes the Universe to accelerate.

\section{Dynamical Stability of Model}\label{sec5}

The dynamical system analysis provides a valuable complement to the cosmological model obtained through the cosmological observations. The constraints used to analyse the  cosmographic parameters as well the dynamics of the Universe mostly at the present stage. However for the comprehensive understanding of the behaviour of the model over a longer period, the dynamical system analysis provides some useful information. Specifically, the stable critical points, which are to be obtained from the dynamical system analysis, provide the behaviour of the model at different evolutionary phases. In addition, a better insight can be achieved on the future evolution of the Universe through the evolutionary behaviour of the density parameters, which are not captured earlier in the paper.
By recasting the cosmological equations in dynamical systems, we can
investigate the stability of the systems in this section \cite{Bahamonde18, Agostino18}. If we take $f(Q,T)=-Q+\Phi (Q,T)$, then Eqs. \eqref{18} and \eqref{19} becomes, 
\begin{widetext}
\begin{eqnarray}
    3H^{2} &=& \kappa\rho-\frac{\Phi}{2}+Q\Phi_{Q}+\frac{2\Phi_{T}}{\kappa+\Phi_{T}}\left[(2Q\Phi_{QQ}+\Phi_{Q}-1)\dot{H}+\Phi_{QT}H\dot{T}\right]~,\label{43}\\
    -2\dot{H}-3H^{2} &=& \kappa p +\frac{\Phi}{2}-Q\Phi_{Q}-2\left[(2Q\Phi_{QQ}+\Phi_{Q})\dot{H}+\Phi_{QT}H\dot{T} \right]~.\label{44}
\end{eqnarray}
\end{widetext}

We consider that the Universe is filled with dust and
radiation fluids, so that, 
\begin{equation}\label{45}
\rho =\rho _{m}+\rho _{r}~,~~~~~~~~~p=\frac{1}{3}\rho _{r}~.
\end{equation}
where $\rho _{r}$ and $\rho _{m}$ are the energy densities of radiation and matter respectively. The Friedmann equations with DE can also be expressed as follows: 
\begin{eqnarray}
3H^{2} &=&\rho +\rho _{de}~,\label{46}\\
2\dot{H}+3H^{2} &=&p+p_{de}.\label{47}
\end{eqnarray}
where $\rho _{de}$ and $p_{de}$ respectively represent the DE density and pressure contribution caused by the geometry. Therefore, geometrical DE- $\Phi (Q,T)$ density and pressure can be expressed as follows:
\begin{widetext}

    \begin{eqnarray}
  \rho_{de} &=& -\frac{\Phi}{2}+Q\Phi_{Q}+\frac{2\Phi_{T}}{\kappa+\Phi_{T}}\left[(2Q\Phi_{QQ}+\Phi_{Q}-1)\dot{H}+\Phi_{QT}H\dot{T}\right]~,\label{48}\\
  p_{de} &=& \frac{\Phi}{2}-Q\Phi_{Q}-2\left[(2Q\Phi_{QQ}+\Phi_{Q})\dot{H}+\Phi_{QT}H\dot{T} \right]~.\label{49}
\end{eqnarray}
\end{widetext}

Now, we may introduce the density parameter pertaining to the
pressureless matter, radiation, and DE respectively as, 
\begin{equation}\label{50}
\Omega _{m}=\frac{\rho _{m}}{3H^{2}},~~~~~~~~~~~~~\Omega _{r}=\frac{\rho _{r}}{3H^{2}},~~~~~~~~~~~~~\Omega _{de}=\frac{\rho _{de}}{3H^{2}}.
\end{equation}

To analyse the dynamics of the cosmological models, we use the following
dimensionless variables: 
\begin{eqnarray}
x &=&\frac{\kappa \rho _{m}}{3H^{2}},~~~~~~~~~~~~y=\frac{\kappa \rho _{r}}{3H^{2}}, \label{51}\\
z &=&\frac{\Phi }{6H^{2}},~~~~~~u=2\Phi _{Q},~~~~~~v=\frac{\Phi _{T}}{\kappa+\Phi _{T}}.\label{52}
\end{eqnarray}

If prime denotes differentiation with respect to the number of e-folds of
the Universe $N=lna$, then using Eqs. \eqref{43}, \eqref{44} and the 
continuity equation, we get the following dynamical system: 
\begin{eqnarray}
x^{\prime } &=&-x\left( 3+2\frac{\dot{H}}{H^{2}}\right) ~,  \label{53} \\
y^{\prime } &=&-2y\left( 2+\frac{\dot{H}}{H^{2}}\right) ~^{\prime }
\label{54} \\
z^{\prime } &=&(u-2z)\frac{\dot{H}}{H^{2}}+\frac{3vx}{2(v-1)}~,
\label{55} \\
u^{\prime } &=&24\dot{H}\Phi _{QQ}-6\rho _{m}\Phi _{QT}~,  \label{56} \\
v^{\prime } &=&\left( \frac{v(1-v)}{\Phi _{T}}\right) (12\dot{H}\Phi
_{TQ}-3\rho _{m}\Phi _{TT})~.  \label{57}
\end{eqnarray}

Also, we have, 
\begin{equation}\label{58}
\frac{\dot{H}}{H^{2}}=\frac{3}{2(2Q\Phi _{QQ}+\Phi _{Q}-1)}\left[ z-u+\frac{y}{3}+1+2\Phi _{QT}\rho _{m}\right] .
\end{equation}

Next we shall analyse the models with the considered forms of $f(Q,T)$ in the problem. So we have, $\Phi (Q,T)=\beta \log \left(\frac{Q}{Q_{0}}\right)+\gamma T$ and 
\begin{eqnarray}
\frac{\dot{H}}{H^{2}} &=&\frac{3u-3z-y-3}{u+2}~, \label{59}\\
\omega _{eff} &=&-1-\frac{2\dot{H}}{3H^{2}}~, \label{60}\\
\omega _{de} &=&-\frac{(\gamma +\kappa )(u(y+9)-6z)}{(u+2)(6\kappa u+\gamma(3u+y+3)-3\kappa z)}~.\label{61}
\end{eqnarray}%
and the dynamical system obtained as, 
\begin{eqnarray}
x^{\prime } &=&\frac{x(2y+6z-9u)}{u+2}~,\label{62}\\
y^{\prime } &=&\frac{2y(-5u+y+3z-1)}{u+2}~,\label{63}\\
z^{\prime } &=&-\frac{(u-2z)(-3u+y+3z+3)}{u+2}-\frac{3vx}{2(v-1)}~, \label{64}\\
u^{\prime } &=&\frac{2u(-3u+y+3z+3)}{u+2}~,\label{65}\\
v^{\prime } &=&0~.\label{66}
\end{eqnarray}

The term $v^{\prime }=0$ further reduces the above system as, 
\begin{eqnarray}
x^{\prime } &=&\frac{x(2y+6z-9u)}{u+2}~,  \label{67} \\
y^{\prime } &=&\frac{2y(-5u+y+3z-1)}{u+2}~,  \label{68} \\
z^{\prime } &=&-\frac{(u-2z)(-3u+y+3z+3)}{u+2}-\frac{3\gamma x}{2\kappa }~,\label{69} \\
u^{\prime } &=&\frac{2u(-3u+y+3z+3)}{u+2}~.  \label{70}
\end{eqnarray}

To determine the dynamical features of the autonomous system, the coupled
equations $x^{\prime }=0$, $y^{\prime }=0$, $z^{\prime }=0$, and $u^{\prime
}=0$ must be solved. The critical points and their description has given in
TABLE -- \ref{tab2} with the stability conditions and the corresponding cosmology. 
\begin{widetext}

\begin{table*}[!htbp]
\begin{center}
\begin{tabular}{c c c c c c c c c c}
\hline 
~~Critical Point~~ & $x$ & $y$ & $z$ & $u$ & $q$ & $\omega_{eff}$ & $\omega_{de}$ & ~~Phase of Universe~~ & ~~Stability~~\\
 \hline\hline
   $A$ & $0$ & $0$ & $\lambda$ & $1+\lambda$ & $-1$ & $-1$ & $-\frac{1}{2+\lambda}$ & ~~DE dominated~~ & \begin{tabular}{@{}c@{}} Stable for \\   $\lambda >-2$ \end{tabular}\\
   $B$ & $0$ & $0$ & $-1$ & $0$ & $-1$ & $-1$ & $-1$ & ~~DE dominated~~ & ~~Stable~~\\
   $C$ & $0$ & $0$ & $0$ & $0$ & $\frac{1}{2}$ & $0$ & $0$ & ~~Matter dominated~~ & ~~Unstable~~\\
   $D$ & $0$ & $1$ & $0$ & $0$ & $1$ & $\frac{1}{3}$ & $0$ & ~~Radiation dominated~~ & ~~Unstable~~\\
   \hline
\end{tabular}
\caption{Critical points for the dynamical system.}
\label{tab2}
\end{center}
\end{table*}

\begin{figure}[h]
\centerline{\includegraphics[scale=0.70]{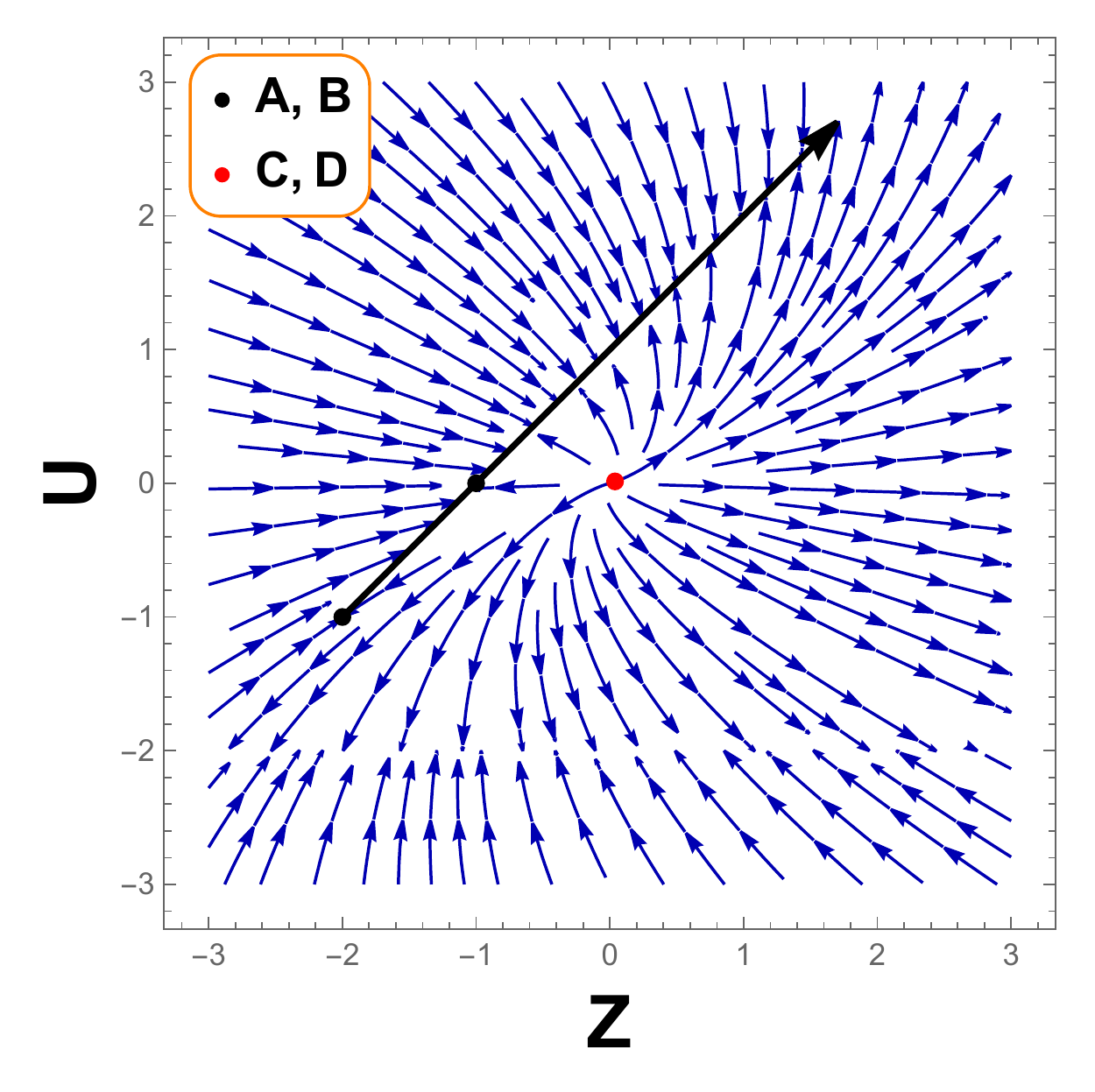}
            \includegraphics[scale=0.70]{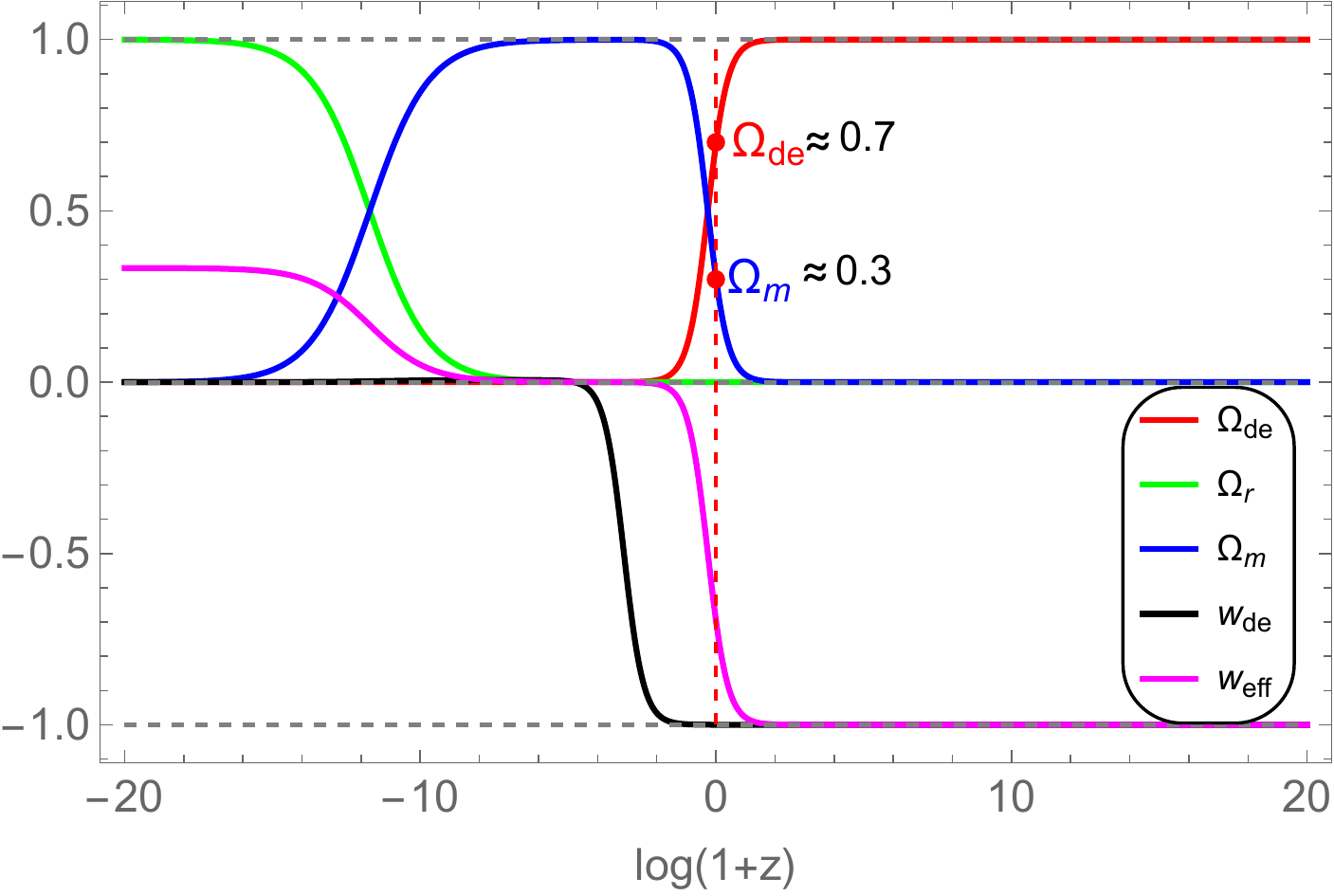}}
    \caption{[\textbf{Left panel}] Phase portrait for the dynamical system in $u-z$ plane. [\textbf{Right panel}] Evolution plot for the considered model using the initial condition $x=10^{-1.9}$, $y=10^{-3}$, $z=10^{-12}$ and $u=10^{-12}$ .}
    \label{fig:16}
\end{figure}
\end{widetext}An important tool for studying dynamical systems is the phase portrait, which displays comparable trajectory plots. The stability of the models can be tested using the phase portrait. FIG. \ref{fig:16} shows the phase portrait for the system given in Eqn. \eqref{67}-\eqref{70}. FIG. \ref{fig:16} (Left Panel) shows the trajectories in the $u-z$ plane. The critical point $A$ and $B$ are stable points and the $C$ and $D$ are unstable once. Also, FIG. \ref{fig:16} (Right Panel) shows the evolutionary behavior of the Universe. At present the $\Omega _{de}\approx 0.7$ and $\Omega _{m}\approx 0.3$, show the accelerated expansion of the Universe.

\section{Conclusion}\label{sec6}
We have presented an accelerating cosmological model of the Universe in $f(Q,T)$ theory of gravity where the parameters are constrained from different cosmological data sets. The $f(Q,T)$ gravity theory leads to the non-conservation of energy-momentum which provides additional force field so that it favours non-geodesic motion of test particles. It should be expected that, the violation of energy-momentum may be responsible for the late phase acceleration of the Universe and as such affects the whole dynamics. In the present work, we have considered  logarithmic form of the functional $f(Q,T)$, which can also be reduced to GR in specific conditions. The free parameters of the Hubble parameter such as $H_{0}$, $\alpha$ and $n$ are constrained using the $Hubble+BAO+SNe$ datasets with the confidence range of $1-\sigma$ and $2-\sigma$. One should expect that, the violation of energy-momentum conservation should enter into the observational survey process and consequently to the process of constraining the model parameters. However, we have not investigated the same in details assuming that the constrained parameters bear the signature of the energy-momentum non-conservation.

The range of best-fit values is provided TABLE -- \ref{tab}. We obtain the present value of the  deceleration parameter as $q_{0}=-0.35^{+0.02}_{-0.02}$ for the $Hubble+BAO+SNe$ datasets. The state finder $(r,~s)$ pair shows the quintessence behavior of the Universe, which has also been validated through the $Om(z)$ diagnostic. 

Further, the dynamical parameters are assessed. The energy density remains positive throughout the evolution and shows a decreasing behavior. The present value of the EoS parameter  obtained as, $\omega_{0}=-0.56^{+0.39}_{-0.39}$ for the $Hubble+BAO+SNe$ datasets. The EoS parameter shows quintessence behavior at present and behaves like $\Lambda$CDM at late time. Also, the energy conditions are analyzed to validate the dynamic behavior of the Universe. The null energy condition shows decreasing behavior and vanishes at late time, whereas, dominated energy condition shows decreasing behavior and remains positive throughout the evolution. The strong energy condition is satisfied at early times and violets at late time, which further satisfies the late time cosmic expansion issue.

Finally, we have analyzed the stability of the accelerating model framed through the logarithmic form of the function $f(Q,T)$ through the dynamical system analysis. Four critical points are obtained and the corresponding cosmological phase has been indicated. We get the stable critical points [$A$ and $B$] and unstable critical points [$C$ and $D$] for the system. The critical points $A$ and $B$ show the DE-dominated phase of the Universe, whereas $C$ shows matter-dominated and $D$ shows radiation dominated phase of the Universe. This has also been confirmed through the phase portrait for the system in the $u-z$ plane. It is worthy to mention here that the points along the straight line (Black) in the phase portrait indicate the stability along the line. The evolution plot also confirms the accelerating behaviour of the Universe and the present value of the density parameter of DE and matter phase are obtained to be $\Omega_{de}\approx 0.7$ and $\Omega_{m}\approx 0.3$ respectively. The behavior of the density parameter for matter in FIG. \ref{fig:16} (Right Panel), shows that it approaches a value of around $0.3$ at present. This value is consistent with the latest observational results from various cosmological probes. It is worth mentioning here that, the MCMC analysis carried out for our model also predicts similar value for the density parameter $\Omega_{m,0}$. In view of this, the analysis presented here supports well the concordance cosmological model in which the matter content of the Universe is dominated by dark matter and baryonic matter with $\Omega_m\approx0.3$. Finally we can conclude that the accelerating behaviour of the model has been confirmed by invoking the cosmological data sets to constrain the free parameters and also from the initial conditions of the dynamical variables of the dynamical system analysis.   \\

\section*{Acknowledgments:} BM acknowledges the support of IUCAA, Pune (India) through the visiting associateship program. The authors are thankful to the anonymous reviewer for the comments and suggestions to improve the quality of the paper.

\textbf{Data availability} There are no new data associated with this
article.

\textbf{Declaration of competing interest} The authors declare that they have no known competing financial interests or personal relationships that could have appeared to influence the work reported in this paper.\newline

\end{document}